\documentclass[epjc3]{svjour3}
\usepackage[english]{babel}
\usepackage[utf8x]{inputenc}

\usepackage{amsmath}
\usepackage[pdftex]{graphicx}
\usepackage{fancyhdr}
\usepackage{amsmath,amssymb}
\usepackage{graphicx}
\usepackage{amssymb}
\usepackage{url,color}
\usepackage{epstopdf}
\usepackage{caption}
\usepackage{subfig}
\usepackage[]{lineno}
\usepackage{anysize}
\usepackage{setspace}
\usepackage{comment} 
\usepackage{lineno}

\newcommand{\ttbar}{ t {\bar t}}
\newcommand{\ttprocess}{e^+ e^- \rightarrow t \bar t}

\newcommand{\eplus}{e^+}
\newcommand{\eminus}{e^-}
\newcommand{\epem}{\eplus\eminus}

\newcommand{\leftp}{\left(}
\newcommand{\rightp}{\right)}

\DeclareMathAlphabet{\mathsfit}{\encodingdefault}{\sfdefault}{m}{sl}

\definecolor{darkred}{rgb}{0.6,0,0}
\definecolor{darkgreen}{rgb}{0,0.6,0}
\definecolor{darkblue}{rgb}{0,0,0.8}
\definecolor{lightyellow}{rgb}{1,1,0.9}
\definecolor{pink}{rgb}{1,0,1}

\def\vev{{\rm v}}

\def\iab{\mbox{ab$^{-1}$}}
\def\ifb{\mbox{fb$^{-1}$}}

\def\TeV{\ifmmode {\mathrm{\ Te\kern -0.1em V}}\else
                   \textrm{Te\kern -0.1em V}\fi}%
\def\GeV{\ifmmode {\mathrm{\ Ge\kern -0.1em V}}\else
                   \textrm{Ge\kern -0.1em V}\fi}%
\def\MeV{\ifmmode {\mathrm{\ Me\kern -0.1em V}}\else
                   \textrm{Me\kern -0.1em V}\fi}%
\def\keV{\ifmmode {\mathrm{\ ke\kern -0.1em V}}\else
                   \textrm{ke\kern -0.1em V}\fi}%
\def\eV{\ifmmode  {\mathrm{\ e\kern -0.1em V}}\else
                   \textrm{e\kern -0.1em V}\fi}%
\let\tev=\TeV
\let\gev=\GeV

\journalname{Eur. Phys. J. C}
\begin{document}
\title{CP-violating top quark couplings at future linear $\epem$ colliders} 


\author{W.~Bernreuther\thanksref{addr1} 
        \and 
        L.~Chen\thanksref{addr1,t1} 
        \and 
        I.~Garc\'{\i}a\thanksref{addr2,t2,t3} 
        \and 
        M.~Perell\'o\thanksref{addr2} 
        \and 
        R.~Poeschl\thanksref{addr3} 
        \and 
        F.~Richard\thanksref{addr3} 
        \and 
        E.~Ros\thanksref{addr2} 
        \and 
        M.~Vos\thanksref{addr2}
}
\thankstext{t1}{Present address: Max-Planck-Institut f. Physik, 80805
M\"unchen, Germany}
\thankstext{t2}{Corresponding author, e-mail: ignacio.garcia@ific.uv.es}
\thankstext{t3}{Present address: CERN, CH-1211 Geneva 23, Switzerland}

\institute{Institut f\"ur Theoretische Teilchenphysik und Kosmologie, RWTH Aachen University, 52056 Aachen, Germany \label{addr1} 
           \and 
           \noindent Instituto de F\'isica Corpuscular (IFIC, UVEG/CSIC), Apartado de Correos 22085, E-46071, Valencia, Spain \label{addr2} 
           \and 
           \noindent Laboratoire de l'Acc\'el\'erateur Lin\'eaire (LAL), Centre Scientifique d'Orsay, 91898 Orsay C\'edex, France~\label{addr3}
}


\date{\today}

\maketitle

\begin{abstract}
We study the potential of future lepton colliders to probe violation of the 
CP symmetry in the top quark sector. In certain extensions of the 
Standard Model, such as the two-Higgs-doublet model (2HDM), sizeable anomalous
top quark dipole moments can arise, that may be revealed by a precise
measurement of top quark pair production. We present results from detailed 
Monte Carlo studies for the ILC at 500~\GeV{} and CLIC at 380~\gev{} and
use parton-level simulations to explore the potential of high-energy 
operation. We find that precise 
measurements in $e^+e^- \rightarrow t\bar{t}$ production with subsequent 
decay to lepton plus jets final states can provide sufficient sensitivity 
to detect Higgs-boson-induced CP violation in a viable two-Higgs-doublet model.
The potential of a linear $e^+e^-$ collider to detect CP-violating electric 
and weak dipole form factors of the top quark exceeds the prospects of the 
HL-LHC by over an order of magnitude.
 \end{abstract}



\section{Introduction}

The top quark is by far the heaviest fundamental particle known to date.
Its large mass implies that it is the Standard Model particle that 
is most strongly coupled to the electroweak symmetry breaking sector. 
The top quark is also set apart 
from the other quarks in that it does not form hadronic bound-states -- 
that is, it offers the possibility 
to study the interactions of a bare quark.
The experimental investigation of single top and top-quark pair production 
at the Tevatron and at the Large Hadron Collider (LHC) has led to a 
precise knowledge of the top-quark strong and weak 
charged-current interactions. These results are in good agreement 
with the Standard Model (SM) predictions.

The electroweak neutral current interactions of the top quark are much 
less precisely investigated. At the LHC an accurate measurement of the 
top-quark neutral current couplings to the photon ($\gamma$) and $Z$ boson is 
challenging, because $\ttbar$ pairs are dominantly produced by the 
strong interactions, and the associated production
of $\ttbar$ and a hard photon or $Z$ boson is relatively rare 
compared to the production of $\ttbar~+~{\rm jets}$.
Future lepton colliders will offer the opportunity to precisely explore 
these top-quark interactions. The studies in 
Refs.~\cite{Amjad:2015mma,Khiem:2015ofa,Janot:2015yza,Devetak:2010na,Rontsch:2015una,Englert:2017dev} 
have shown that linear collider (LC) experiments can measure the top-quark 
electroweak couplings with very competitive precision. 

Several projects exist for $e^+e^-$ colliders with sufficient energy to 
produce top-quark pairs (i.e., with centre-of-mass energy $\sqrt{s} > 2 m_t$).
A mature design exists for a linear $e^+e^-$ collider that
can ultimately reach centre-of-mass energies up to approximately 
1~\tev, the International Linear Collider (ILC)~\cite{Baer:2013cma}, 
to be hosted in Japan. We focus on the planned operation
at a centre-of-mass energy of 500~\gev{}, consider the initial 
integrated luminosity scenario (500~\ifb{}) and the nominal 
H20 scenario~\cite{Barklow:2015tja} (4~\iab{}).

Extensive R\&D into high-gradient acceleration has moreover 
opened up the possibility of a relatively compact multi-\tev{} 
collider, the Compact Linear Collider (CLIC)~\cite{Linssen:2012hp}. 
The CLIC program envisages an initial stage 
that collects approximately
500~\ifb{} at $\sqrt{s} =$ 380~\gev, followed by operation at a 
centre-of-mass energy of up to 3~\tev~\cite{clicstaging}.

Both linear collider projects offer the possibility of polarized beams.
Operation of the collider with two polarization configurations
allows to disentangle the photon and $Z$-boson form factors.
ILC and CLIC both envisage polarizing the electron beam 
(80\% longitudinal polarization). The ILC baseline design envisages
30\% positron polarization. In the CLIC design this is left as
an upgrade option. 

ILC and CLIC have developed detailed detector designs and sophisticated 
simulation and reconstruction software, which allows a careful study
of experimental effects in realistic conditions. Here, we perform a 
full simulation of the reaction $e^+e^-\to \ttbar \to$ lepton plus jets 
in the context of these projects.

In this paper we extend the studies of 
Refs.~\cite{Amjad:2015mma,Amjad:2013tlv} to couplings that violate the
combination of charge conjugation and parity (CP, in the following).
New physics that affects top-quark 
production and/or decay is parametrized in terms of form factors that 
depend on kinematic invariants. The Standard Model predicts 
CP violation in top-quark pair production and decay to be very small, 
well beyond the experimental sensitivity of existing and planned facilities
(see the discussion in Section~\ref{sec:systematics}).
Some extensions of the SM, such as, for instance, two-Higgs-doublet models, 
can give rise to sizeable effects~\cite{Atwood:2000tu}.
Any observation of CP violation in the top-quark sector would be 
clear evidence of physics beyond the SM. Early studies of non-standard CP 
violation (i.e. CP violation that is not induced by the Kobayashi-Maskawa
CP phase) in $e^+e^- \to t{\bar t}$ include those in 
Refs.~\cite{Kane:1991bg,Atwood:1991ka,bib:cpvbernreuther0}. 
Our  study is based on the observables proposed in 
Ref.~\cite{bib:cpvbernreuther2} for the lepton plus jets final states that
have a direct sensitivity to the electric and weak dipole form factors of the
top quark. We present the first result of a detailed Monte Carlo simulation
of these observables in a realistic experimental environment.

This paper is organized as follows. Section~\ref{sec:anomalous} describes 
our conventions for the form factors
with which one can parametrize top-quark pair production at lepton colliders.
We analyze in Section~\ref{sec:cpvff} the potential magnitude of the CP-violating top-quark electric and 
weak dipole form factors in two SM extensions taking into account present experimental constraints.
Moreover, we briefly discuss the potential magnitude of
CP-violating form factors in top-quark decay, $t\to W b$.
Observables and associated asymmetries sensitive to CP violation in 
$\ttbar$ production which apply to lepton plus jets final states 
are introduced in Section~\ref{sec:observables}. 
We study the effect of polarized beams in
Section~\ref{sec:polarized} and determine the relations between the CP 
asymmetries and CP-violating form factors.
Full simulation results at two centre-of-mass energies are presented 
in Section~\ref{sec:CPILC} for the ILC at 500~\gev{} and 
in Section~\ref{sec:CPCLIC} for CLIC at 380~\gev. In 
Section~\ref{sec:CPCLIC3000} we study the prospect of 1-3~\tev{} operation
in a parton-level study.
Systematic uncertainties are discussed in Section~\ref{sec:systematics}.
The prospects of linear $e^+e^-$ colliders for the extraction of the
CP-violating form factors derived in this study are presented in 
Section~\ref{sec:precision} and compared with other studies in 
the literature of the potential of lepton and hadron colliders.
We conclude in Section~\ref{sec:conclusions}. 


\section{CP violation in $e^+e^-\to t{\bar t}$}
\label{sec:anomalous}

  Our present knowledge about physics at the TeV scale implies that 
  in $e^+e^-$ collisions at centre-of-mass (c.m.) energies $\lesssim 1$ TeV 
  top-quark pairs will be dominantly produced by SM interactions,
  to wit, by s-channel photon and $Z$-boson exchange.  
  New physics interactions that involve the top quark may modify  
  the $\ttbar X$ $(X=\gamma,Z)$ vertices and 
  the overall $e^+e^-\to t{\bar t}$ production amplitude. 
 In order to pursue a relatively model-independent analysis, we
  assume that new CP-violating interactions, which can lead to sizeable
 effects in $t{\bar t}$ production, have only a small effect on
top-quark decay. We will discuss in Section~\ref{sec:cpvff} the validity
of this assumption within two SM extensions.

Lorentz covariance determines the structure of the $t\bar{t} X$ vertex. 
In the case that both top quarks are on their mass shell and the photon 
and $Z-$boson are off-shell we can write the $t\bar{t} X$ vertex as: 
\begin{equation}
\Gamma^{ttX}_{\mu}(k^2) = -ie\left\{\gamma_{\mu}\left( F_{1V}^X(k^2) 
+\gamma_5 F_{1A}^X(k^2) \right)   + \frac{\sigma_{\mu\nu}k^{\nu}}{2m_t} 
 \left(  iF_{2V}^X(k^2) +\gamma_5 F_{2A}^X(k^2)  \right) \right\},
\label{eq:vtxvtt}
\end{equation}
where $e=\sqrt{4\pi\alpha}$, with $\alpha$ the electromagnetic fine structure 
constant, $m_t$ denotes the mass of the top quark, and $k^\mu=q^\mu + \bar{q}^\mu$ is 
the sum of the four momenta $q^\mu$ and $\bar{q}^\mu$ of the $t$ and $\bar{t}$~quark. We
 use $\gamma_5=i\gamma^0\gamma^1\gamma^2\gamma^3$ and   
 $\sigma_{\mu\nu}=\frac{i}{2}\leftp \gamma_{\mu}\gamma_{\nu} - \gamma_{\nu}\gamma_{\mu} \rightp$. The ${F}_{i}$ denote form factors which are to be probed 
in the time-like domain $k^2 > 4 m_t^2$ by the reaction at hand\footnote{The form factors ${F}_{i}$ are related to the $\tilde{F}_{i}$ of Ref.~\cite{bib:snow2005} through the following relations: 
\begin{equation}
\label{eq:rel1}
\widetilde F^X_{1V} = -\left( F^X_{1V}+F^X_{2V} \right) \, , \qquad
\widetilde F^X_{2V}  =  F^X_{2V} \, , \qquad
\widetilde F^X_{1A} = -F^X_{1A} \, , \qquad
\widetilde F^X_{2A} =  -iF^X_{2A} \, .
\end{equation}
.}.

For off-shell  $\gamma, Z$ bosons, there are in general two more contributions, one of which could violate CP invariance. However, if 
   the mass of the electron is neglected, an excellent approximation 
in our case, these two terms will not  contribute to the $t{\bar t}$ 
production amplitude. We therefore omit them in the following.

Within the Standard Model, and at tree level, the $F_1$  are related to the 
electric charge of the top quark $Q_t$ and its weak isospin:
\begin{equation}
F_{1V}^{\gamma,SM}= Q_{t} = -\frac{2}{3},\,F_{1A}^{\gamma,SM}=0,\,F_{1V}^{Z,SM}=
 -\frac{1}{4s_W c_W}\left(1-\frac{8}{3}s^2_W \right),\,F_{1A}^{Z,SM}=\frac{1}{4s_W c_W},
\label{eq:ffactors}
\end{equation}
where $s_W$ and $c_W$ are the sine and the cosine of the weak mixing angle $\theta_W$. 
The chirality-flipping form factors $F_2$ are zero at tree level. As in any
 renormalizable theory they must be loop-induced. At zero momentum transfer
$F^{\gamma}_{2V}(0)$ is related via $F^{\gamma}_{2V}(0) =Q_{t}(g_t-2)/2$ to 
the anomalous magnetic moment of the top quark $g_t$, where $Q_t$ denotes its electrical charge in units  of $e$.

In this paper we focus on the form factors $F_{2A}^X$ that violate the 
combined charge and parity symmetry CP. The electric dipole moment of the top quark 
 is determined by $F_{2A}^\gamma$ for an on-shell photon at zero momentum transfer,
 $d_t^\gamma= -(e/2m_t)F_{2A}^\gamma(0)$. In analogy to this relation 
 one may define an electric dipole
 form factor (EDF) and  a weak dipole form factor (WDF) for on-shell $t, {\bar t}$ but
  off-shell $\gamma, Z$:
  \begin{equation}
    d_t^X(k^2) = - \frac{e}{2 m_t} F_{2A}^X(k^2) \,, \quad X=\gamma, Z.
   \label{eq:EDFWDF} 
  \end{equation}
For off-shell gauge bosons these form factors are in general gauge-dependent.
However, within the two SM extensions that will be discussed in the next section, the $d_t^X(s)$
are gauge-invariant to one-loop approximation. This may justify their use in parametrizing possible CP-violating 
 effects in $t{\bar t}$ production.
 
 Finally we note that new physics effects are often described in the framework of effective field theory (EFT)
 by anomalous couplings, i.e., constants. The `couplings' $d_t^\gamma$ and $d_t^Z$ can be related to the 
coefficients of certain dimension-six operators, cf., for instance, Ref.\cite{AguilarSaavedra:2008zc}. 
However, by using EFT for describing new physics one assumes that there is a gap between 
   the typical energy scale of the process under consideration ($\sqrt{s}$ in our case) and the scale of new physics. 
This is not the case for the models
   that we consider in the next section. In particular, in the kinematic domain that we are interested in, $d_t^\gamma$ and $d_t^Z$ show a non-negligible dependence on $\sqrt{s}$ and can develop absorptive parts, therefore becoming complex.


 \section{CP-violation in SM extensions}
 \label{sec:cpvff}
  In the SM, where CP violation is induced by the  Kobayashi-Maskawa (KM) phase in the charged weak current
  interactions, resulting CP effects in flavour-diagonal amplitudes are too small to be measurable in 
   $e^+e^-\to t{\bar t}$~\cite{Atwood:2000tu}. 
   Sizeable CP-violating effects involving top quarks may arise in SM extensions with additional, non-KM CP-violating interactions.
   In this section we consider two extensions of this type, namely two-Higgs-doublet models and the mimimal supersymmetric extension 
    of the Standard Model (MSSM), and assess the potential magnitude of the top-quark EDF and WDF in these models, taking into account present experimental 
    constraints. At the end of this section we briefly discuss  the potential size of CP-violating  form factors 
     in top-quark decay, $t\to W b$.
  \subsection{Two-Higgs-doublet models}
  \label{suse:2hdm}
   In view of its large mass the top quark is an excellent probe of non-standard CP violation generated by an extended Higgs sector.
 We consider here two-Higgs-doublet models (2HDM) where the SM is extended by an additional Higgs doublet field
and where the Yukawa couplings of the Higgs doublets $\Phi_1, \Phi_2$ are such that no tree-level flavour-changing neutral currents (FCNC) are
 present. The physical Higgs boson spectrum of these models consists of a charged Higgs boson and its antiparticle, $H^\pm$, and three
  neutral Higgs bosons, one of which is to be identified with the 125 GeV Higgs resonance. The Higgs potential $V(\Phi_1, \Phi_2)$ can violate 
  CP, either explicitly or spontaneously by Higgs fields developing a vacuum 
expectation value with non-trivial CP-violating phase. If this is the case, 
then the physical CP-even and CP-odd neutral Higgs fields mix. 
  In the unitary gauge the resulting three neutral Higgs mass eigenstates $h_j$ are related to the two neutral CP-even states $h$, $H$,
   and the CP-odd state $A$  by an orthogonal transformation:
   \begin{equation}
    (h_1,h_2,h_3)^T = R~(h, H, A)^T \, .
    \label{eq:orthotr}
   \end{equation}
The orthogonal matrix $R$ is parametrized by three Euler angles\footnote{We use  the conventions of \cite{Bernreuther:2015fts}.}
 that are related to the parameters of the Higgs potential. 

  For phenomenological studies it is useful to choose as independent parameters of the 2HDM a set 
that includes the masses $m_j$ and $m_+$ of the three neutral and the charged Higgs boson, respectively, the three Euler angles $\alpha_i$ of $R$,
 the parameter $\tan\beta=v_2/v_1$ which is the ratio of the vacuum expectation values of the two Higgs-doublet fields, and 
 ${\vev}=\sqrt{v_1^2 + v_2^2}$ which is fixed by experiment to the value 
 ${1/\vev} =(\sqrt{2}G_F)^{1/2} = 246$ GeV. In case of CP violation in the Higgs sector, the Higgs mass eigenstates $h_j$ couple to both scalar and 
 pseudoscalar fermion currents. The Yukawa Lagrangian is
 \begin{equation} \label{eq:LfZ}
  {\cal L}_Y = -\frac{m_f}{\vev}\left( a_{jf} {\bar f}f
    - b_{jf}{\bar f} i\gamma_5 f\right) h_j  \, .
  \end{equation}  
Here $f$ denotes a quark or charged lepton and the reduced scalar and pseudoscalar Yukawa couplings  $a_{jf}, b_{jf}$
 depend on the type of 2HDM \cite{Branco:2011iw}, on the matrix elements of $R$, and on $\tan\beta$.

Within the 2HDM, the CP-violating part of the scattering amplitude of  $e^+e^-\to t{\bar t}$ is determined (to one-loop approximation and in the limit of vanishing electron mass $m_e$) entirely by the top-quark EDF and WDF  \eqref{eq:EDFWDF} 
  that are induced at one-loop by CP-violating neutral Higgs boson exchange \cite{Bernreuther:1992dz}. There are no CP-violating box contributions.
 (A CP-violating scalar form factor $\widetilde{F}_{S}^Z(s)$ is also generated, but it does not contribute for $m_e=0$.) Thus the one-loop
 top-quark EDF and WDF generated in 2HDM are gauge invariant.

The real and imaginary parts of the top-quark EDF $d^\gamma_t(s)$ and WDF $d^Z_t(s)$ were computed for several types of 2HDM in  \cite{Bernreuther:1992dz}.
The EDF  $d_t^\gamma(s)$ is generated at one loop by the CP-violating exchange of  the three Higgs bosons $h_j$ between 
the outgoing $t$ and $\bar t$. A $CP$-violating Higgs potential implies that the
$h_j$ are not mass-degenerate. The form factor becomes complex, i.e., it has an absorptive part for $s>4 m_t^2$. We remark that 
 $d_t^\gamma(s)\propto m_t^3$: two powers of $m_t$ result from the Yukawa interactions  \eqref{eq:LfZ} and one power from the necessary chirality flip.
The one-loop WDF  $d_t^Z(s)$ receives two different contributions: the first one is topologically identical to   $d_t^\gamma(s)$, but with the tree-level 
$t{\bar t}$ coupling to the photon replaced by the vectorial $t{\bar t}$ coupling to the $Z$ boson. The second one involves the 
$ZZh_j$ coupling (where only the CP-even component of $h_j$ is coupled) and the pseudoscalar coupling of  $h_j$ to the top quark.
The second contribution, which is proportional to $m_Z^2m_t$, becomes complex for $s>(m_Z+m_j)^2$, where $m_j$ is the mass of $h_j$.

  Before evaluating the formulae for the top-quark EDF and WDF given in \cite{Bernreuther:1992dz} we  discuss present experimental constraints 
  on the parameters of the type-II 2HDM. The 125 GeV Higgs resonance must be identified with one of the neutral Higgs bosons $h_j$.
  For definiteness, we identify it with $h_1$ and assume the other two neutral Higgs bosons to be heavier.
  The ATLAS and CMS results on the production and decay  of the 125 GeV Higgs resonance $h_1(125~{\rm GeV})$ 
  imply that this boson is Standard-Model like;
  its couplings to weak gauge bosons, to the $t$ and $b$ quark, and to $\tau$ leptons have been determined with a precision of $10 - 25~\%$~\cite{Khachatryan:2014jba,Aad:2015gba,Khachatryan:2016vau} and these results are in reasonable agreement with the SM predictions. 
  Moreover, the investigation of angular correlations in $h_1(125~{\rm GeV})\to Z Z^*\to 4 \ell$ exclude
  that this Higgs boson is a pure pseudoscalar $(J^P=0^-)$ \cite{Chatrchyan:2012jja,Aad:2013xqa}. However, 
  this does not imply that $h_1$ is purely CP-even $(J^P=0^+)$ -- it can be a CP mixture. 
   Because the pseudoscalar component of such a state does not 
   couple to $ZZ$ and $WW$ at tree level, a potential pseudoscalar 
component is difficult to detect in the decays of $h_1$ to 
   weak bosons\footnote{The decays 
$h_1\to\tau\tau$, where a CP-violating effect
occurs at tree level if $h_1$ is a CP mixture, may be used to check
   whether or not $\phi_1$ has a pseudoscalar component. See, for
instance \cite{Berge:2008wi}. Other possibilities include the
associated production of $t{\bar t} h_1$, once a sufficiently large
event sample will have been collected.}.
   
   In the following we assume that $h_1$ is a CP-mixture with couplings to fermions and gauge bosons that are in accord with the 
   LHC constraints \cite{Khachatryan:2014jba,Aad:2015gba}. We are interested in 2HDM parameter scenarios where the couplings of 
   the $h_j$ to top quarks are not suppressed as compared to the corresponding SM Yukawa coupling. This is the case for 
   $\tan\beta \sim 1$ or somewhat lower than one. Moreover, we assume that the two other neutral Higgs bosons $h_2$, $h_3$
   are significantly heavier than $h_1$.  In the type-II 2HDM and for $\tan\beta < 1$ the Yukawa couplings of the $h_j$ to $b$ quarks and $\tau$ leptons 
   are suppressed as compared  to the corresponding SM Yukawa couplings, cf. Table~\ref{tab:numfWZ}.
    Moreover, the constraint that $h_1$ has SM-like couplings to weak gauge bosons implies that the couplings of $h_2$, $h_3$ to
    $WW$ and $ZZ$ are small, irrespective of the CP nature of these Higgs bosons. This follows from a sum rule, cf., for instance, \cite{Branco:2011iw}.
    2HDM parameter scenarios with  $\tan\beta \lesssim 1$ and 
    $h_2$, $h_3$ masses equal or larger than about 500 GeV are compatible with the non-observation of heavy neutral Higgs bosons at the 
    LHC in final states with electroweak gauge bosons~\cite{ATLAS:2013nma,Khachatryan:2015lba,Aad:2014yja,Aad:2015wra,Khachatryan:2015cwa},
    $b$ quarks~\cite{Nagai:2013xwa,Khachatryan:2015yea}, charged leptons~\cite{Khachatryan:2014wca,Aad:2014vgg}, 
    and top quarks~\cite{ATLAS:2016pyq}.
    The charged Higgs boson $H^\pm$ of 2HDM is of no concern to us here. Constraints from $B$-physics data, in particular 
    from the rare decays $B\to X_s+\gamma$ and $B^0 -{\bar B}^0$ mixing imply that the mass of $H^\pm$ must be 
    larger than $\sim 700$ GeV  for low values of $\tan\beta$ \cite{Hermann:2012fc}. 
   
   In order to assess the potential size of the form factors $F_{2A}^X(s)$ $(X=\gamma,Z)$ in type-II
    2HDM with Higgs sector CP violation we make a scan over the  independent parameters that are of relevance for this analysis. 
    In the kinematic range $\sqrt{s}\lesssim 500$ GeV the most important contribution to the top-quark EDF and WDF arise from $h_1$ exchange if this
     Higgs boson has top-quark Yukawa couplings such that the modulus of the product $a_{1t}b_{1t}$ is about one. 
     We take into account recent constraints on the couplings of $h_1$ to $W,Z, t, b,\tau$ \cite{Khachatryan:2016vau} and the experimental constraints on the masses and couplings of $h_2, h_3$ from Ref.~\cite{Khachatryan:2014jba,Aad:2015gba,ATLAS:2013nma,Khachatryan:2015lba,Aad:2014yja,Aad:2015wra,Khachatryan:2015cwa,Nagai:2013xwa,Khachatryan:2015yea,Khachatryan:2014wca,Aad:2014vgg,ATLAS:2016pyq}. 
      We vary $\tan\beta$ in the range $0.35 \leq \tan\beta \leq 1$ and the three Higgs mixing angles in the range $-\pi/2\leq \alpha_i\leq \pi/2$,
      determine the resulting reduced Yukawa couplings $a_{jf}, b_{jf}$ and the couplings $f_{jVV}$ of the $h_j$ to $ZZ$. 
      A benchmark set of resulting couplings is given in Table~\ref{tab:numfWZ}. Somewhat tighter constraints on the CP-violating top-Higgs couplings were derived in Ref.~\cite{Kobakhidze:2016mfx}.

\vspace{2mm}
\begin{table}[htbp]
\begin{center}
  \caption{Benchmark values of the reduced couplings of the neutral Higgs bosons $h_j$ to quarks, leptons, and weak gauge bosons
   with $|a_{1t}b_{1t}| \gtrsim 1$  that are in accord with present experimental constraints. The couplings $f_{jVV}$ to the weak gauge bosons are given in units of $m_Z^2/v$.}
  \vspace{1mm}
\begin{tabular}{c|cccccc}
             & $a_{jt}$  & $a_{jb} = a_{j\tau}$  & $b_{jt}$  & $b_{jb} = b_{j\tau}$  & $f_{jVV}$ \\ \hline 
   $h_1$ & $1.379$  &  $0.881$ & $0.910$   & $0.111$ &   $0.935$\\
   $h_2$ &  $-0.569$ &   $-0.275$ & $2.706$  & $0.331$ &  $-0.307$\\
   $h_3$ &  $-2.634$ & $0.521$ & $-0.108$  & $0.484$  & $-0.013$ \\ \hline
\end{tabular}
\label{tab:numfWZ} 
\end{center}
\end{table}
       For calculating  the form factors  $F_{2A}^X(s)$ we assume that the Higgs bosons $h_2$ and $h_3$ are heavier than 500 GeV.
       For definiteness we set their masses to be 1200 GeV and 600 GeV, respectively.
       Using the formulae of Ref.~\cite{Bernreuther:1992dz}, $m_t=173$ GeV, and the values of
        the Higgs couplings of  Table~\ref{tab:numfWZ}, the real and imaginary parts of $F_{2A}^X(s)$ 
        are shown as  functions of the c.m. energy in  Fig.~\ref{fig:ewdf}. In the kinematic range displayed in these plots the
         imaginary  part of the EDF is about three times larger than that of the WDF. This holds true also for the real parts of the form factors close 
          to the $\ttbar$ threshold, while they become significantly smaller in magnitude around $\sqrt{s}=500$ GeV due to the strong fall-off of the dominant 
          contribution from $h_1$. 
          The values of the real and imaginary parts of these
   form factors are listed in Table~\ref{tab:numFF} for two c.m. energies that are chosen for the simulations in  Sec.~\ref{sec:polarized} - \ref{sec:CPCLIC}.
  
 \begin{figure}[h!]
 \begin{center}
 {\includegraphics[width=0.48\textwidth]{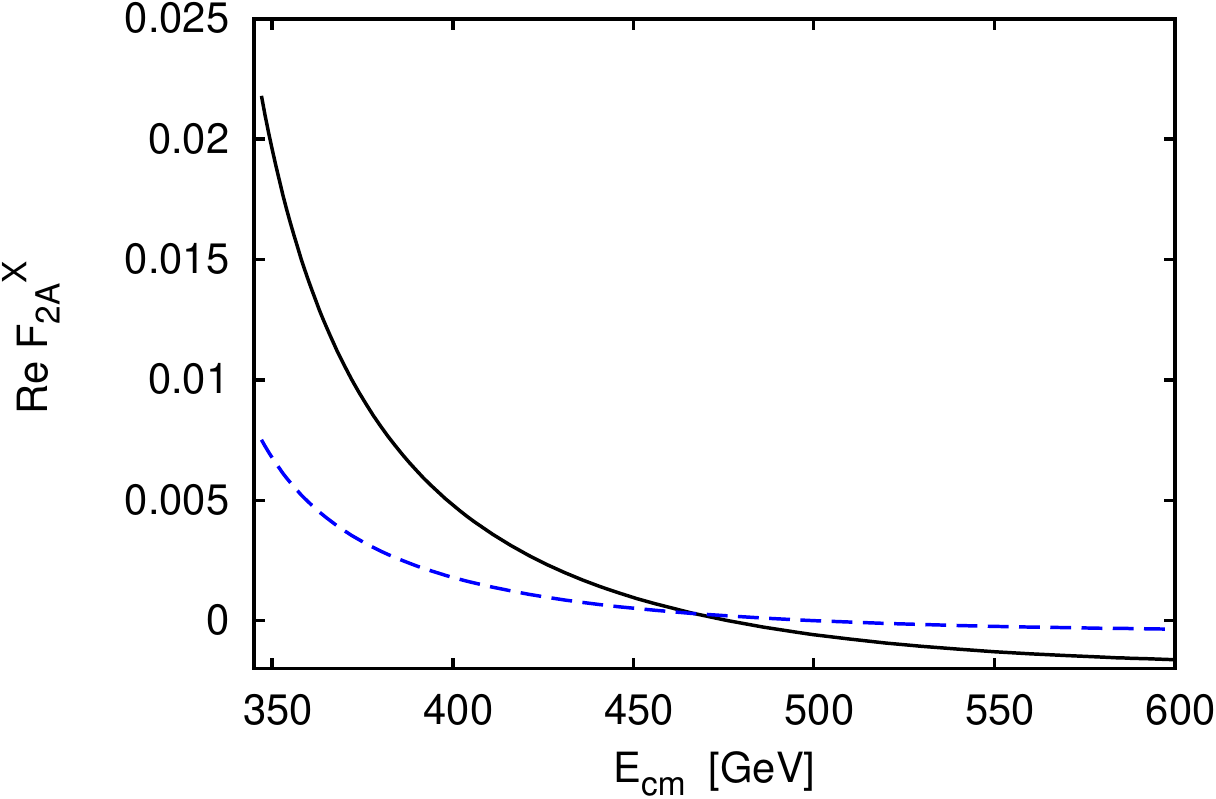}}	
 {\includegraphics[width=0.48\textwidth]{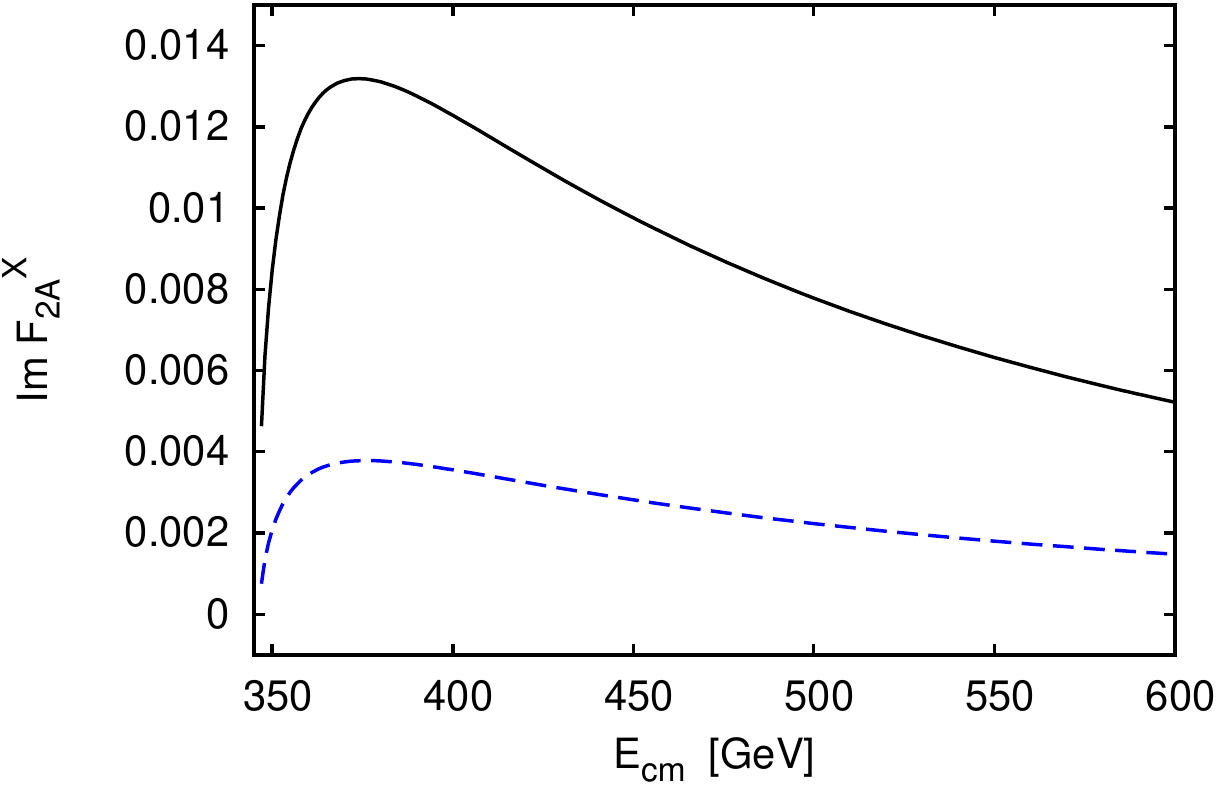}}
 \caption{Left panel: The real part of the top-quark EDF $F_{2A}^\gamma$ (solid, black) and WDF  $F_{2A}^Z$ (dashed, blue),
  evaluated with the couplings of Table~\ref{tab:numfWZ} and neutral Higgs-boson masses
  $m_1=125$ GeV, $m_2=1200$ GeV, and $m_3=600$ GeV, as a function of the c.m. energy.
  Right panel: The same for the imaginary part of the top-quark EDF and WDF.
 }
 \label{fig:ewdf}
 \end{center}
\end{figure}

 \vspace{2mm}
\begin{table}[h!]
\begin{center}
  \caption{Values of the real and imaginary parts of the top-quark EDF and WDF for two c.m. energies. Input parameters are as in 
  Fig.~\ref{fig:ewdf}.}
  \vspace{1mm}
\begin{tabular}{c|cccc}
    $\sqrt{s}$ [GeV] &  ${\rm Re}~F_{2A}^\gamma$ &  ${\rm Re}~F_{2A}^Z$ & ${\rm Im}~F_{2A}^\gamma$ & ${\rm Im}~F_{2A}^Z$ \\ \hline 
   380 &  $8.1\times 10^{-3}$ & $2.9\times 10^{-3}$ & $1.3\times 10^{-2}$ & $3.8\times 10^{-3}$ \\
    500  & $-0.6\times 10^{-3}$ & $0.7\times 10^{-6}$ & $7.8\times 10^{-3}$ & $2.2\times 10^{-3}$ \\ \hline
\end{tabular}
\label{tab:numFF} 
\end{center}
\end{table}      

       In the kinematic range that we are interested in ($\sqrt{s} \lesssim 500$ GeV) the imaginary parts of the EDF and WDF are rather insensitive
        to the values of the heavy Higgs-boson masses, as long as $m_{2,3}> 500$ GeV. This is also the case for the real parts of the form factors close to 
        the $\ttbar$ threshold that are dominated by the contribution from $h_1$ exchange. This term falls off strongly with increasing 
        c.m. energy. Moreover, at c.m. energies $\sqrt{s}\gtrsim 500$ GeV the contributions from $h_2, h_3$ to the real parts of the form factors may 
        no longer be negligible. We find that the real parts of the  EDF and WDF at  $\sqrt{s} = 500$ GeV depend, for fixed Higgs-boson couplings, 
        sensitively on the masses of $h_2, h_3$, but do not exceed $10^{-3}$ in magnitude for the couplings of Table~\ref{tab:numfWZ}.

    As mentioned above, the formulae of  \cite{Bernreuther:1992dz} apply to any type of 2HDM where  tree-level  FCNC are absent.
     In fact, the results shown in  Fig.~\ref{fig:ewdf} and given in Table~\ref{tab:numFF} apply also to other types of 2HDM in the low 
      $\tan\beta$ region; for instance, to the type-I model where all right-chiral quarks and charged leptons are coupled to the Higgs doublet
       $\Phi_2$ only,  or to the so-called lepton specific model
     where the right-chiral quarks (right-chiral charged leptons) are coupled to $\Phi_2$ $(\Phi_1)$ only.
         
    In summary,  within the 2HDM the real (imaginary) part of the top-quark electric dipole form factor  $F_{2A}^\gamma$ can be as large as $\sim 0.02$ ($\sim 0.01$) in magnitude near the  $\ttbar$ production threshold, taking into account the present constraints from LHC data.

  \subsection{The minimal supersymmetric SM extension}
  \label{suse:mssm}
  
  The Higgs sector of the MSSM corresponds to a type-II 2HDM. Supersymmetry (SUSY) forces the tree-level Higgs potential  $V(\Phi_1, \Phi_2)$
  of the MSSM to conserve CP.  Nevertheless, the MSSM contains in its general form many CP-violating phases besides the KM phase, especially 
  in the supersymmetry-breaking terms of the model, including phases of the complex Majorana mass terms of the neutral gauginos
  and of the complex chargino and sfermion mass matrices. Motivated by assumptions about SUSY breaking at very high energies, one 
  often puts constraints on the SUSY-breaking terms, in particular on the CP-violating phases, in order to restrict the number of unknown 
  parameters of the model. Nevertheless, generic features of SUSY CP violation remain. 
   Unlike the case of Higgs-boson induced one-loop EDMs, fermion EDMs generated at one-loop can be large, also for $u$, $d$ quarks and the electron.
  The experimental upper bounds on the EDM of the neutron and of atoms/molecules strongly constrain
  in particular the CP-phases associated with the sfermion mass matrices of the first and second generation, barring fine-tuned cancellations. See, for instance, Ref.~\cite{Pospelov:2005pr} for a review.  However, the phases 
  of the sfermion mass matrices need not be flavour-universal. 
   For the top flavour the associated phase $\varphi_{\tilde t}$
   can still be of order one. Often a common phase of the gaugino masses is assumed. Using phase redefinitions of the fields in the MSSM Lagrangian,
   one can choose for the parametrization  of MSSM CP violation in the top-quark sector~\cite{Bartl:1997iq,Hollik:1998vz}
    the phase $\varphi_{\tilde t}$, the 
    corresponding  $b$-flavour phase $\varphi_{\tilde b}$, and the phase $\varphi_\mu={\rm arg}(\mu)$ of the so-called $\mu$ term in the MSSM Lagrangian 
     that generates a Dirac mass of the higgsinos. For a rather recent  analysis of constraints on the  CP-violating phases in the MSSM, see Ref.~\cite{Arbey:2014msa}.

   The one-loop top-quark EDF and WDF induced by the CP-violating interactions of the  MSSM are gauge invariant.
   They are generated by one-loop $\gamma\ttbar$ and $Z\ttbar$ vertex diagrams involving ${\tilde t}$ and ${\tilde b}$ squarks,  gluinos ${\tilde g}$,
   neutralinos ${\tilde\chi^0}$, and charginos ${\tilde \chi}^\pm$ in the loop.
   The ${\tilde t}{\tilde t}^* {\tilde g}$ contributions to the EDF and WDF were determined in 
   \cite{Christova:1992ee,Bernreuther:1993xp,Grzadkowski:1993kb}.
   The complete set of 1-loop contributions  were computed in \cite{Bartl:1995gj,Bartl:1997iq,Hollik:1998vz,Hollik:1998wk}.
   They consist, apart from the gluino contribution, of the chargino contribution (with ${\tilde \chi}^+{\tilde \chi}^-{\tilde b}$
    and ${\tilde b}{\tilde b}^* {\tilde \chi}^+$ in the loop), and of the neutralino contribution (with 
    ${\tilde t}{\tilde t}^* {\tilde \chi}^0$ and ${\tilde \chi}^0 {\tilde \chi}^0 {\tilde t}$ in the loop).

     If light neutralinos and charginos and/or  light ${\tilde t}$, ${\tilde b}$  squarks
     with masses $m_i, m_j$ of order $100 - 200$ GeV would exist there would be strong enhancements
     of the top-quark EDF and WDF $F_{2A}^{\gamma,Z}(s)$ in the range $2 m_t \lesssim \sqrt{s} \lesssim 500$ GeV 
     near the two-particle production threshold $\sqrt{s}_{th}=m_i + m_j$. Refs. \cite{Bartl:1997iq,Hollik:1998vz}
     computed these form factors for light gauginos and  ${\tilde t}, {\tilde b}$ squarks. Ref.~\cite{Hollik:1998vz} found 
     maximal values of the form factors at $\sqrt{s}=500$ GeV for some favorable set of SUSY parameters
     of the order of $10^{-3}$.
         However, the input parameters of these computations have since been excluded.
     Searches for supersymmetry were negative so far, and the 
     LHC searches put strong lower bounds on the masses of SUSY particles that are, in most cases, model-dependent, to wit:
     $m_{\tilde g} > 1.8$ TeV, $m_{{\tilde b}_{1,2}}>840$ GeV, $m_{{\tilde\chi}^\pm}> 715$ GeV, and
     $m_{{\tilde t}_{1,2}}>800$ GeV for $m_{{\tilde \chi_1}^0} < 200$ GeV. For a recent review, we refer to Ref.\cite{AdamICHEP}. 
However, a light stop with mass $\sim 200$ GeV is not yet excluded if the decay ${\tilde t}_1\to t {\tilde \chi_1}^0$ exists. The 
limit  $m_{{\tilde\chi}^\pm}> 715$ GeV results from an analysis in Ref.~\cite{Aad:2014nua} using a simplified SUSY model.
    
      In order to estimate the potential size of the top-quark EDF and WDF, we evaluated the chargino, gluino, and neutralino
       contributions using the formulae of \cite{Bartl:1997iq,Hollik:1998vz}
            with SUSY masses that are in accord with these experimental constraints. The phases 
       $\varphi_\mu,\varphi_{\tilde t},\varphi_{\tilde b}$  were chosen such that they maximize the EDF and WDF for given masses. 
Using the lower bounds on the masses of SUSY particles cited in the previous paragraph we find: 
       \begin{equation} \label{eq:susyres}
        |{\rm Re}~F_{2A}^\gamma|, \; |{\rm Re}~F_{2A}^Z| < 10^{-3} \, ,\quad   |{\rm Im}~F_{2A}^\gamma|,\; |{\rm Im}~F_{2A}^Z| < 10^{-4} \quad 
         \text{for} \; \sqrt{s} \lesssim 500~{\rm GeV.}
       \end{equation}
       As mentioned above, a light top squark ${\tilde t}_1$ with mass $\sim 200$ GeV and also a  light neutralino ${\tilde \chi_1}^0$
        is not yet excluded. In this case non-zero but small imaginary parts are generated by the gluino and neutralino contribution to the EDF and WDF
         in the considered range of c.m. energies.

  In the case of the MSSM there are also  CP-violating box contributions to $\ttprocess$ that involve 
  neutralino (${\tilde e} {\tilde \chi}^0 {\tilde \chi}^0 {\tilde t}$)  and chargino 
  (${\tilde \nu}_e {\tilde \chi}^\pm {\tilde \chi}^\mp {\tilde b}$)  exchanges
   in the one-loop amplitudes. 
   They are, as shown in \cite{Hollik:1998vz,Hollik:1998wk}, in general not negligible compared to the top-quark EDF and WDF contributions.
   We shall, however, refrain from evaluating these box contributions, which goes beyond the scope of this paper. 
   In the simulations performed in  Sec.~\ref{sec:polarized} - \ref{sec:CPCLIC} we shall stick to the parametrization \eqref{eq:vtxvtt}
   of CP-violating effects in $\ttbar$ production in terms of the EDF and WDF.

  \subsection{CP-violating form factors in $t\to W b$}
  \label{suse:twb}
So far, the only top-quark decay mode that has been observed is $t\to W b$ 
with subsequent decay of the $W$ boson into leptons or quarks.
In the SM the branching ratio of this decay is almost 100 percent. 
 The decay amplitude for $t \to W^+ b$ with all particles on-shell 
can be parametrized in terms of two chirality-conserving and two 
 chirality-flipping form factors $f_L, f_R$ and $g_L, g_R$, 
respectively; cf., for instance, \cite{bib:cpvbernreuther0}.
 The measurements of these form factors~\cite{Khachatryan:2014vma,Aad:2015yem} 
are in agreement with the SM predictions.
 
 Let us  denote the corresponding form factors in the charge-conjugate 
decay ${\bar t}\to W^- {\bar b}$ by $f'_i, g'_i$ $(i=L,R)$.
 CPT invariance implies that $f^*_i =f'_i$ and  $g^*_i = g'_i$. CP invariance 
requires that the corresponding form factors are equal.
 These relations imply the following: if final-state interactions can be 
neglected in top-quark decay, then CP violation induces non-zero
 imaginary parts that are equal in magnitude but differ in 
sign~\cite{bib:cpvbernreuther0,Bernreuther:2008us}:
 ${\rm Im} f'_i=-{\rm Im} f_i$,  ${\rm Im} g'_i=-{\rm Im} g_i$, $i=L,R$.
 
 In Ref.~\cite{Bernreuther:2008us} the potential size of CP-violating 
(and CP-conserving) contributions to the form factors in $t\to W b$ was 
investigated for several SM extensions. Within the 2HDM it was found 
that $|{\rm Im} f_i|,|{\rm Im} g_i| \lesssim 3\times 10^{-4}$ 
  for $\tan\beta\gtrsim 0.6$. In the MSSM the CP-violating effects were 
found to be smaller by at least one order of magnitude.
 The observables and CP-violating asymmetries that we introduce
  in the next section and in section~\ref{sec:polarized} are insensitive 
to CP violation in top-quark  decay. 
  Therefore we  can neglect CP violation in top-quark decay in the following and parametrize CP violation in $\ttbar$ production with subsequent 
   decay into lepton plus jets final states solely by the top-quark 
EDF and WDF  defined in Eq.~\eqref{eq:vtxvtt}. One may probe CP violation in 
 semi-leptonic $t$ and $\bar t$ decay with a CP-odd asymmetry constructed from
  suitable triple product correlations \cite{bib:cpvbernreuther0,Bernreuther:1993xp}.

  \subsection{Synopsis}
  \label{suse:synops}
  Let us summarize the discussion of the previous subsections. 
We analyzed the potential size of CP-violating  effects in $\ttbar$ 
production in $e^+e^-$ collisions and subsequent $t$ and $\bar t$ decay 
within two popular and motivated SM extensions, taking into account present 
experimental constraints.
 As to the BSM scenarios  investigated above, an extended Higgs sector with 
CP-violating neutral Higgs boson exchange has the largest potential to 
generate observable effects in this reaction. If the observed 
$h_1(125{\rm GeV})$ Higgs resonance has both scalar and pseudoscalar 
couplings to top quarks whose strengths are of order one compared to 
the SM top Yukawa coupling then the magnitude of ${\rm Im}~F_{2A}^\gamma$ 
can be $\sim 1\%$ in the energy range $\sqrt{s}\lesssim 500$ GeV that 
we consider in the following. The real part of this form factor can become 
of the same order of magnitude near the $\ttbar$ threshold. The real and 
imaginary parts of the top-quark WDF are in general smaller
   by a factor of about 0.3, cf. Table~\ref{tab:numFF}. Within the MSSM 
the top-quark EDF and WDF are smaller, with maximum values compatible
with current experimental constraints below $10^{-3}$. The CP-violating 
form factors in the $t \to Wb$ decay amplitude that can be generated 
within the 2HDM or the MSSM are very small and of no further interest to 
us here.
   Moreover, we recall that within the 2HDM  there are no CP-violating box 
contributions to the $e^+e^-\to\ttbar$ amplitude to one-loop approximation 
if the electron mass is neglected.  These results motivate the use of the 
parametrization of Eq.~\eqref{eq:vtxvtt} in the simulations of the 
following sections.


\section{Optimal CP-odd observables}
\label{sec:observables}

As demonstrated in Ref.~\cite{Amjad:2015mma}, at a future linear $e^+e^-$ 
collider precise measurements of 
the $\ttbar$ cross-section and the top-quark forward-backward 
asymmetry for two different beam polarizations allow the extraction of 
the top-quark CP-conserving electroweak form factors with a precision 
that exceeds that of the HL-LHC. In this section the prospects for 
the measurement of CP-violating 
form factors $F_{2A}^{\gamma,Z}$ are investigated, as an extension of the 
previous study. The CP-violating effects in $e^+e^- \rightarrow t\bar{t}$ 
manifest themselves in specific top-spin effects, namely CP-odd top 
spin-momentum correlations and $t\bar{t}$ spin correlations. 
If one considers the dileptonic decay channels, 
$t\bar{t}\to\ell^+ \ell'^- + ...$, then it is appropriate to consider CP-odd 
 dileptonic angular correlations~\cite{bib:cpvbernreuther0}, which efficiently 
trace CP-odd $t\bar{t}$ spin correlations. We recall the well-known fact 
that the charged lepton  in semi-leptonic $t$ or $\bar t$ decay is by far 
the best analyzer of the top spin. Here we consider  $t\bar{t}$ decay to 
lepton plus jets final states which yield more events than the dileptonic 
channels and, moreover, allow for a straightforward experimental 
reconstruction of the $t$ and $\bar t$ rest frames. For these final states 
the most efficient way to probe for CP-violating effects in $t\bar{t}$ 
production is to construct observables that result from $t$ and ${\bar t}$ 
single-spin momentum correlations, that is, from correlations which involve 
only the spin of the semi-leptonically decaying $t$ or ${\bar t}$. 
Here, we adopt the  observables proposed in~\cite{bib:cpvbernreuther2} for 
detecting these correlations in lepton plus jets final states. 

We consider  in the following the production of a top-quark pair via the 
collision of longitudinally polarized electron and positron beams:
	
\begin{equation}
 e^+ ({\bf p}_+,P_{e^+})  +  e^- ({\bf p}_-, P_{e^-})  \quad \to \quad t( {\bf k}_t) +  \bar{t} ({\bf k}_{\bar{t}}) \, .
\label{eq:TheReaction}
\end{equation} 
Here, ${\bf p}_\pm$ and ${\bf k}_t, {\bf k}_{\bar{t}}$ denote the $e^\pm$, $t$, 
and ${\bar t}$ three-momenta in the $e^+e^-$ c.m. frame.
The spin degrees of freedom of the $t$ and $\bar t$ are not exhibited. 
Moreover, $P_{e^-}$ $(P_{e^+})$ is the longitudinal polarization  degree of 
the electron (positron) beam. In our notation, $P_{e^-}=-1$ $(P_{e^+}=-1)$  
refers to left-handed electrons (positrons). 
For our purpose the most useful final states are, as mentioned,  
the lepton plus jets final states from semi-leptonic $t$ decay and 
hadronic $\bar{t}$ decay and vice versa:
\begin{equation}
t  \enskip  \bar{t}  \quad \to \quad \ell^+({\bf q}_+) + \nu_\ell + b + \overline{ X}_{\rm had}({\bf q}_{\bar X})\, ,
\label{eq:TopDecay}
\end{equation}
\begin{equation}
  t  \enskip  \bar{t} \quad \to \quad  X_{\rm had}({\bf q}_X) + \ell^-({\bf q}_-) +  \bar \nu_\ell     + \bar{b} \, ,
\label{eq:AntiTopDecay}
\end{equation}
 where the three-momenta in \eqref{eq:TopDecay} and \eqref{eq:AntiTopDecay} 
 also refer to the $e^+ e^-$ c.m. frame. 

We compute the reactions \eqref{eq:TheReaction} - \eqref{eq:AntiTopDecay} at 
tree level, both in the SM
 and with non-zero CP-odd form factors $F_{2A}^{\gamma,Z}$,
taking the polarizations and spin correlations of the 
intermediate $t$ and ${\bar t}$ into account.
 As discussed in the previous section these form factors can 
have imaginary  parts. Non-zero real parts ${\rm Re} F_{2A}^{\gamma,Z}(s)$ 
induce a difference in the $t$ and $\bar{t}$ polarizations 
 orthogonal to the scattering plane of the reaction. Non-zero 
absorptive parts, ${\rm Im} F_{2A}^{\gamma,Z}(s)$, lead to a difference 
in the $t$ and $\bar{t}$ polarizations along the top-quark direction of 
flight and along the direction of the electron or positron beam. 
At the level of the intermediate  $t$ and ${\bar t}$ these effects manifest 
themselves in non-zero expectation values of the following CP-odd observables:
\begin{equation} 
 \left(\hat{\bf p}_+ \times  \hat{\bf k}_t\right)\cdot ({\bf s}_t - {\bf s}_{\bar t}) \, , \quad
  \hat{\bf k}_t \cdot  ({\bf s}_t - {\bf s}_{\bar t} ) \, , \quad 
  \hat{\bf p}_+\cdot  ({\bf s}_t - {\bf s}_{\bar t} ) \, ,
\label{eq:expvaltt}
\end{equation}
where ${\bf s}_t$ and ${\bf s}_{\bar t}$ denote the spin operators 
of $t$ and ${\bar t}$, respectively and hats denote unit vectors.
In \eqref{eq:expvaltt} two-body kinematics is used, i.e., 
${\bf k}_{\bar t} = -  {\bf k}_t$.
 The expectation value of the first observable 
of the list~\eqref{eq:expvaltt} depends on ${\rm Re} F_{2A}^{\gamma,Z}$, 
while the expectation values of the other two observables depend on 
 ${\rm Im} F_{2A}^{\gamma,Z}$. Each observable listed in \eqref{eq:expvaltt} 
is the difference of two terms that involve 
  the $t$ and $\bar t$ spin, respectively. The term that contains the 
$t$ $({\bar t})$ spin can be translated, in the case of the lepton plus jets
   final states, into a correlation that involves the $\ell^+$ $(\ell^-)$ 
direction of flight. This is the most efficient way 
to analyze the $t$ $({\bar t})$ spin. These correlations can be 
measured with the  $\ell^+$ + jets and $\ell^-$ + jets 
events \eqref{eq:TopDecay} and \eqref{eq:AntiTopDecay}, respectively.

Based on these considerations,  so-called optimal 
observables~\cite{Atwood:1991ka}, i.e., observables 
with a maximal signal-to-noise ratio to a certain parameter appearing
 in the squared matrix element, were constructed in 
Ref.~\cite{bib:cpvbernreuther2} for tracing 
CP violation in the lepton plus jets  final states \eqref{eq:TopDecay} 
and \eqref{eq:AntiTopDecay}. These optimal observables are, in essence, 
given by those parts 
of the squared matrix element that are linear in the CP-violating form 
factors ${\rm Re} F_{2A}^{\gamma,Z}$ or ${\rm Im} F_{2A}^{\gamma,Z}$. 
 One may simplify these expressions and use for the final 
states \eqref{eq:TopDecay} the following two 
observables~\cite{bib:cpvbernreuther2} that are nearly optimal:  
\begin{eqnarray}
  \mathcal{O}_{+}^{Re} & = & (\hat{\bf q}_{\bar{X}} \times \hat{\bf q}_{+}^{*})\cdot \hat{\bf p}_+ \, , \label{eq:ORe} \\
  \mathcal{O}_{+}^{Im}& = & -[1+(\frac{\sqrt{s}}{2m_t}-1)(\hat{\bf q}_{\bar{X}} \cdot \hat{\bf p}_{+})^2]\hat{\bf q}_{+}^{*}\cdot\hat{\bf q}_{\bar{X}}
   +\frac{\sqrt{s}}{2m_t}\hat{\bf q}_{\bar{X}}\cdot\hat{\bf p}_{+}\hat{\bf q}_{+}^{*} \cdot\hat{\bf p}_{+} \, .
\label{eq:OIm} 
\end{eqnarray}

The corresponding observables $\mathcal{O}_{-}$ for the final 
states \eqref{eq:AntiTopDecay} are defined to be the CP image of 
$\mathcal{O}_{+}$ and are obtained from $\mathcal{O}_{+}$ by the substitutions 
$\hat{\bf q}_{\bar{X}} \rightarrow -\hat{\bf q}_{X}$, $\hat{\bf q}_{+}^{*} \rightarrow -\hat{\bf q}_{-}^{*}$, $\hat{\bf p}_{+}\rightarrow\hat{\bf p}_{+}$.   
 The unit vectors $\hat{\bf q}_{\pm}^{*}$ refer  to the $\ell^\pm$ directions of flight defined 
in the $t$ and $\bar t$  rest frame, respectively. The differences of the 
expectation values of $\mathcal{O}_{+}$ and $\mathcal{O}_{-}$ that we consider 
in the next section probe for CP-violating effects.

The observables \eqref{eq:ORe} and \eqref{eq:OIm} are approximations 
to the rather unwieldy optimal observables listed in the 
appendix of Ref.~\cite{bib:cpvbernreuther2}. Using the optimal observables 
at low energy leads to a minor increase in sensitivity. 
Between the $t{\bar t}$ production threshold and $\sqrt{s}\sim$ 500~\gev{} 
the sensitivity to the CP-odd form factors increases by a few percent.
At very high energy the difference is somewhat more pronounced: at 3~\tev{}
the sensitivity  is expected to increase by approximately 30 percent.

As discussed in Section~\ref{suse:twb}, non-standard CP-violating 
interactions can induce, besides CP violation in $t{\bar t}$ production, 
also anomalous couplings in the $t\to W^+b$ and  ${\bar t}\to W^-{\bar b}$ 
decay amplitudes. 
 However, observables such as \eqref{eq:ORe} and \eqref{eq:OIm} and their 
CP images, where the $t$ and $\bar t$ spins are analyzed  by charged lepton 
angular correlations, are insensitive to these anomalous couplings, 
  as long as one uses the linear approximation~\cite{Bernreuther:1993xp,Rindani:2000jg,Grzadkowski:2002gt} which is legitimate here. This justifies the 
parametrization of the CP asymmetries
  $\langle \mathcal{O}_{+} \rangle -\langle \mathcal{O}_{-} \rangle$ solely 
in terms of  $F_{2A}^{\gamma,Z}$.

\section{Polarized beams}
\label{sec:polarized}

We study the distributions of $\mathcal{O}_{-}^{Re}$ and 
$\mathcal{O}_{-}^{Im}$ at leading-order (LO) in the SM couplings, 
putting \mbox{$F_{2A}^{\gamma,Z}=0$}, with the  WHIZARD 1.95 event 
generator~\cite{bib:whizard}. Distributions of both
observables are shown in Fig.~\ref{fig:observables} for a centre-of-mass energy
of 500~\gev. The three histograms in each panel correspond to unpolarized
beams (dashed line), to a left-handed electron beam and a 
right-handed positron beam ($e_L^- e_R^+$, $P_{e^-},P_{e^+} = -80\%, +30\%$,
red continuous histogram)
and for a right-handed electron beam and a left-handed
positron beam ($e_R^- e_L^+$, $P_{e^-},P_{e^+} = +80\%, -30\%$, 
black continuous histogram). The degree
of longitudinal polarization that is used follows the design values of 
the ILC: $P_{e^-},P_{e^+} = \pm 80\%, \mp 30\%$.  
  As the top-quark EDF and WDF are negligible in the SM (and set to
zero in the simulation), the distributions 
for unpolarized beams are symmetric around the origin.

\begin{figure}[h!]
	{\centering
		{\includegraphics[width=0.48\textwidth]{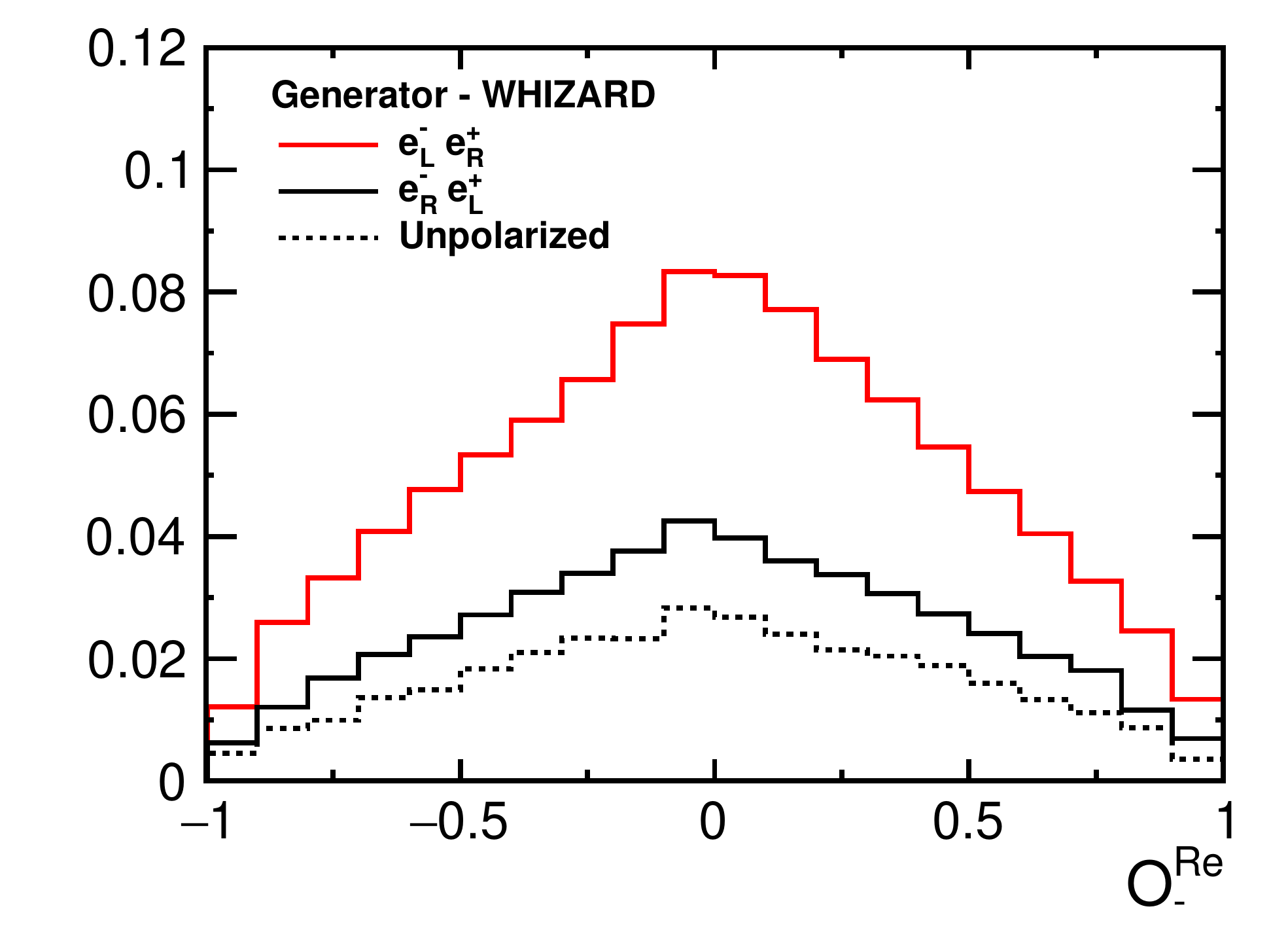}}	
		{\includegraphics[width=0.48\textwidth]{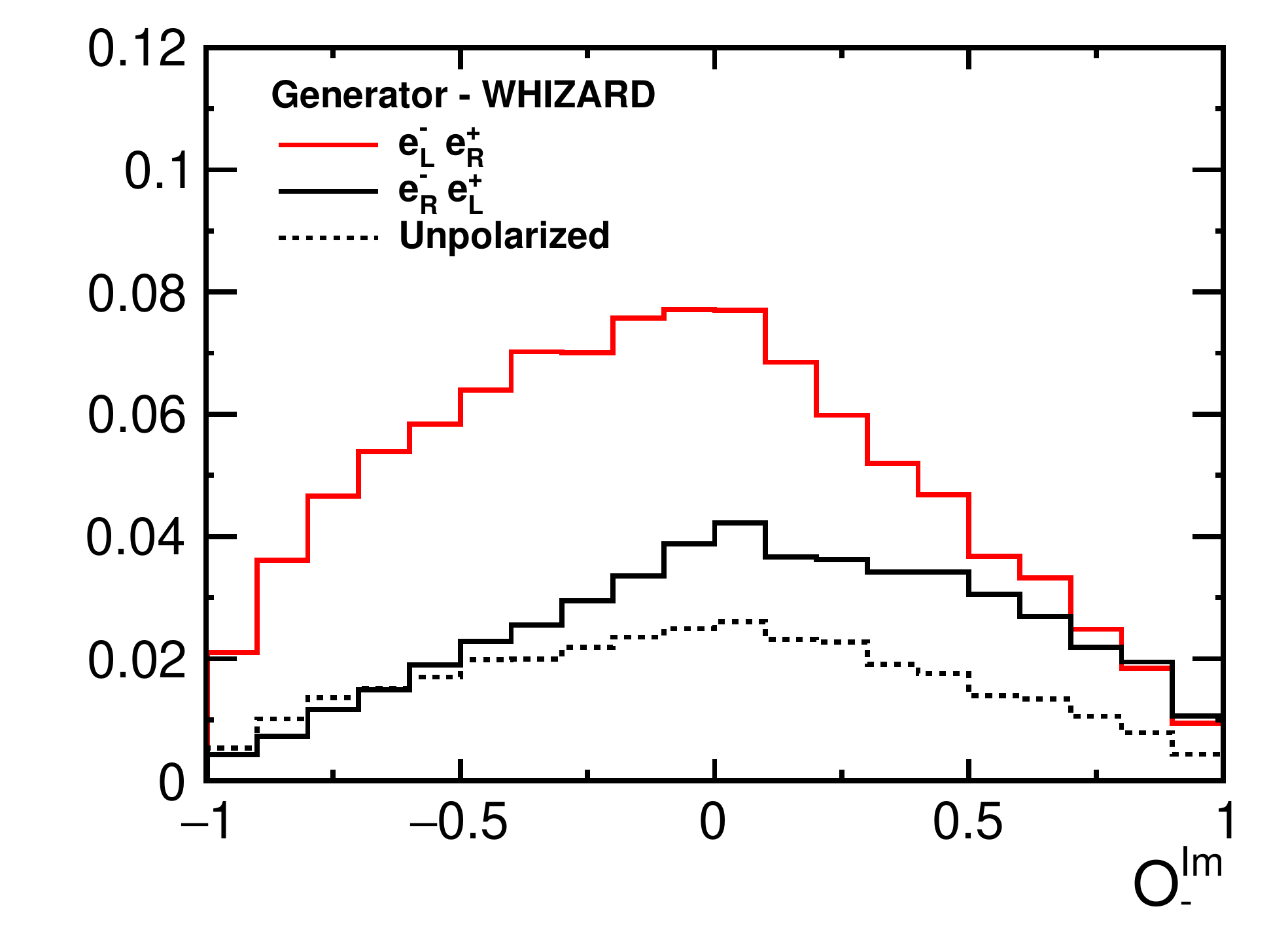}}
	\caption{The WHIZARD LO Standard Model prediction for the 
normalized distribution of the CP observables $\mathcal{O}_{-}^{Re}$ 
(left panel) and $\mathcal{O}_{-}^{Im}$ (right panel) 
defined in \eqref{eq:ORe} and 
\eqref{eq:OIm}. The results correspond to $e^+ e^-$ collisions at a 
centre-of-mass energy of 500~\gev{} and three different beam polarizations: 
the dashed line corresponds to unpolarized beams, the red (black) solid lines 
to -80\% (+80\%) polarization of the electron beam and +30\% (-30\%) 
polarization of the positron beam. The histogram for LR polarized beams
is normalized to unit area. The area of the other histograms 
is scaled so as to maintain the cross section ratios. 
The $\mathcal{O}_{\pm}^{Re}$
distribution is confined to [-1,1] by construction, the $\mathcal{O}_{\pm}^{Im}$ 
distribution is truncated to the same interval.
\label{fig:observables}	
}
}
\end{figure}

Initial-state polarization affects the normalization, but leaves the shape of 
the $\mathcal{O}_{-}^{Re}$ distribution unaffected. The total cross section 
increases strongly for $e^+e^-$ beams in the $e_L^- e_R^+$ configuration as compared to
 unpolarized beams, and somewhat less strongly for the polarization configuration $e_R^- e_L^+$.

Beam polarization has a more profound impact on $\mathcal{O}_{-}^{Im}$ as 
shown in Fig.~\ref{fig:observables} (right panel). With unpolarized beams 
the distribution is symmetric around zero, but the distributions 
corresponding to polarized beams show significant distortions. This is 
expected because the initial state with different beam polarization for 
electrons and positrons is not CP-symmetric. 

Asymmetries $\mathcal{A}$ can be defined~\cite{bib:cpvbernreuther2} 
as the difference of the expectation values
$\langle \mathcal{O}_{+} \rangle$ and $\langle \mathcal{O}_{-} \rangle$: 
\begin{equation}
  \mathcal{A} = \langle\mathcal{O}_+(s,\hat{\bf q}_{+}^{*},\hat{\bf q}_{\bar{X}},\hat{\bf p}_{+})\rangle
  - \langle\mathcal{O} _-(s,\hat{\bf q}_{-}^{*},\hat{\bf q}_{X},\hat{\bf p}_{+})\rangle \, .
  \label{eq:asymmetry}
  \end{equation} 

In the asymmetry, many experimental effects are expected to cancel.
This applies also to the distortion of the $\mathcal{O}_{\pm}^{Im}$
distributions by beam polarization. The $\mathcal{O}_{+}^{Im}$ and 
$\mathcal{O}_{-}^{Im}$ distributions are shifted by approximately 
equal amounts, but in opposite directions. The mean value of the 
$\mathcal{O}_{-}^{Im}$ observable is -0.08$\pm$0.01 
for $P_{e^-},P_{e^+} = -80\%,+30\%$ and +0.09$\pm$0.01 for 
$P_{e^-},P_{e^+} = +80\%,-30\%$. The distributions of $\mathcal{O}_{+}^{Im}$ 
are distorted in the same way as those of $\mathcal{O}_{-}^{Im}$. Therefore, 
the effect of initial-state polarization cancels in the difference of 
both observables.
 
The asymmetries $\mathcal{A}^{Re},\mathcal{A}^{Im}$ are sensitive to CP 
violation effects in the $t\bar{t}$ production amplitude through the 
contributions of ${\rm Re} F_{2A}^{\gamma,Z}$ and ${\rm Im} F_{2A}^{\gamma,Z}$, 
respectively:
%
%
%


 \begin{equation}
 \mathcal{A}^{Re} =\langle \mathcal{O}_{+}^{Re} \rangle  - \langle \mathcal{O}_{-}^{Re} \rangle 
 = c_{\gamma}(s) {\rm Re} F_{2A}^{\gamma} + c_Z(s) {\rm Re} F_{2A}^{Z}  \, , 
 \label{eq:ARe}
\end{equation}
\begin{equation}
  \mathcal{A}^{Im} =\langle \mathcal{O}_{+}^{Im} \rangle - \langle \mathcal{O}_{-}^{Im} \rangle 
  = {\tilde c}_{\gamma}(s) {\rm Im} F_{2A}^{\gamma}   + {\tilde c}_{Z}(s) {\rm Im} F_{2A}^{Z} \, . 
\label{eq:AIm}
\end{equation}
 The values of these coefficients depend on the polarizations $P_{e^-}$ and 
$P_{e^+}.$ In our approach, where we normalize the expectation values 
$\langle \mathcal{O} \rangle$ by the SM cross section (that is, neglecting 
the contributions bilinear in the CP-violation form factors), the asymmetries  
$\mathcal{A}^{Re},  \mathcal{A}^{Im}$ are strictly linear in the form factors. 
Analytical expressions for the coefficients $c_{\gamma}(s), c_{Z}(s), {\tilde c}_{\gamma}(s)$ and ${\tilde c}_{Z}(s)$ of relations~\eqref{eq:ARe} and~\eqref{eq:AIm} 
for arbitrary beam polarization are given in the Appendix. Values for 100\%
polarization are given in Tables~\ref{tab:coeff-pol-LR} 
and~\ref{tab:coeff-pol-RL}, using $m_t=173.34$ GeV, $m_Z = 91.1876$ GeV,  
  $ m_W=80.385$ GeV, and $\sin^2\theta_W =1-m_W^2/m_Z^2$.

\begin{table}[htb!]
  \begin{center}
   \caption{The values of the coefficients in the expressions for the 
asymmetries  ${\mathcal A}^{Re}$ and ${\mathcal A}^{Im}$. The values are calculated for several c.m. energies used in this paper and for 
the $e_L^- e_R^+$ beam polarization configuration ($P_{e^-}=-1$, $P_{e^+}=+1$). } 
    \begin{tabular}{c|cc|cc}
      \hline
     c.m. energy $\sqrt{s}$ [GeV] & $c_{\gamma}(s)$ & $c_{Z}(s)$ &  ${\tilde c}_
{\gamma}(s)$ &  ${\tilde c}_{Z}(s)$ \\ \hline
     380  &   0.245 &       0.173 &     0.232 &      0.164  \\
     500  &   0.607 &      0.418 &      0.512 &      0.352 \\   
     1000 &   1.714 &      1.151 &      1.464 &      0.983 \\
     1400 &   2.514 &       1.681 &     2.528 &      1.691 \\
     3000 &   5.589 &       3.725 &    10.190 &      6.791  \\ \hline
    \end{tabular}
    \label{tab:coeff-pol-LR}
  \end{center} 
\end{table}

\begin{table}[htb!]
  \begin{center}
   \caption{Same as Table~\ref{tab:coeff-pol-LR}, but for the opposite $e_R^- e_L^+$ beam polarization: $P_{e^-}=+1$, $P_{e^+}
=-1$.  } 
    \begin{tabular}{c|cc|cc}
      \hline
     c.m. energy $\sqrt{s}$ [GeV] & $c_{\gamma}(s)$ & $c_{Z}(s)$ &  ${\tilde c}_
{\gamma}(s)$ &  ${\tilde c}_{Z}(s)$ \\ \hline
     380  &   -0.381 &       0.217 &   0.362  &   -0.206 \\
     500  &   -0.903 &       0.500 &   0.761  &   -0.422 \\
     1000 &   -2.437 &       1.316 &   2.081  &   -1.124 \\
     1400 &   -3.549 &       1.909 &   3.569  &   -1.920 \\
     3000 &   -7.845 &       4.205 &  14.302 &    -7.667 \\ \hline
    \end{tabular}
    \label{tab:coeff-pol-RL}
  \end{center} 
\end{table}

The  polarization of the $e^-$ and $e^+$ beams provides a means to 
disentangle the contributions  of the CP-violating photon and $Z$-boson 
vertices. The coefficients $c_{\gamma}(s)$ and ${\tilde c}_{Z}(s)$ 
corresponding to the LR and RL configurations have opposite signs. 
The measurement of the two CP
asymmetries $\mathcal{A}^{Re}$ and $\mathcal{A}^{Im}$ for
two beam polarizations provides sufficient constraints to solve
the system of equations formed by Eq.~\eqref{eq:ARe} and~\eqref{eq:AIm}.

For $\sqrt{s}\gg 2m_t$ the coefficients  $c_{\gamma}(s), c_{Z}(s)$  that appear 
in the expression for $\mathcal{A}^{Re}$ grow with the c.m. energy 
$\sqrt{s}$. The interactions associated with 
$F_{2A}^{\gamma,Z}$ involve a factor $k^\nu$, which is the sum of the $t$ and 
$\bar t$ four-momenta (cf. Eq.~\eqref{eq:vtxvtt}). Therefore, the sensitivity
of the asymmetry $\mathcal{A}^{Re}$ to $F_{2A}$
increases with centre-of-mass energy.

The observables 
$\mathcal{O}_{\pm}^{Im}$ consist of a sum of terms, two of which contain 
the factor $\sqrt{s}$. Therefore the coefficients 
${\tilde c}_{\gamma}(s), {\tilde c}_{Z}(s)$ that determine  
$\mathcal{A}^{Im}$ grow with $s$ for $\sqrt{s}\gg 2m_t$.
However this does not imply that this asymmetry has a significantly higher 
sensitivity than  $\mathcal{A}^{Re}$ to CP-violating effects in $t{\bar t}$ 
production at high energies, because the widths of the distributions of  
$\mathcal{O}_{\pm}^{Im}$ grow accordingly.

\section{Full simulation: ILC at 500~\gev}
\label{sec:CPILC}

In this section we study the 500~\gev{} run of the ILC, assuming
an integrated luminosity of 500~\ifb{}. The sample is divided into two 
beam-polarization configurations: the $LR$ sample has -80\% and +30\%  
electron and positron polarization, respectively.
In the $RL$ sample the signs of both electron and positron polarization are 
inverted: the electron polarization is +80\% and the positron 
polarization is -30\%. 

The full-simulation study is based on samples produced for the 
ILC TDR~\cite{Behnke:2013lya}.
The event sample is generated with WHIZARD 1.95~\cite{Kilian:2007gr}
by the LCC generator group.  
It includes all six-fermion processes that produce a 
lepton plus jets final state, $e^+e^- \rightarrow b\bar{b}l^{\pm}\nu_l q\bar{q}$.
This includes top-quark pair production and a number of other processes 
that lead to the same final state, 
with the largest non-doubly-resonant contribution coming from single top 
production~\cite{Fuster:2015jva}. The effect of Initial-State-Radiation (ISR) 
is included in the generator. Events are generated with the nominal 
ILC luminosity spectrum described in Ref.~\cite{Behnke:2013lya}, which
includes the effects of beam energy spread and beamstrahlung. 
The events generated are restricted to the physics of the SM, hence 
the $F_{2A}^{\gamma,Z}$ are set to zero. Fragmentation
and hadronization is modelled using PYTHIA 6.4~\cite{Sjostrand:2006za}
 with a parameter set tuned to $e^+e^-$ data recorded
at LEP.

\begin{figure}[h!]
	{\centering
	\subfloat[\small $\mathcal{O}_{+}^{Re}$]{\includegraphics[width=0.5\textwidth]{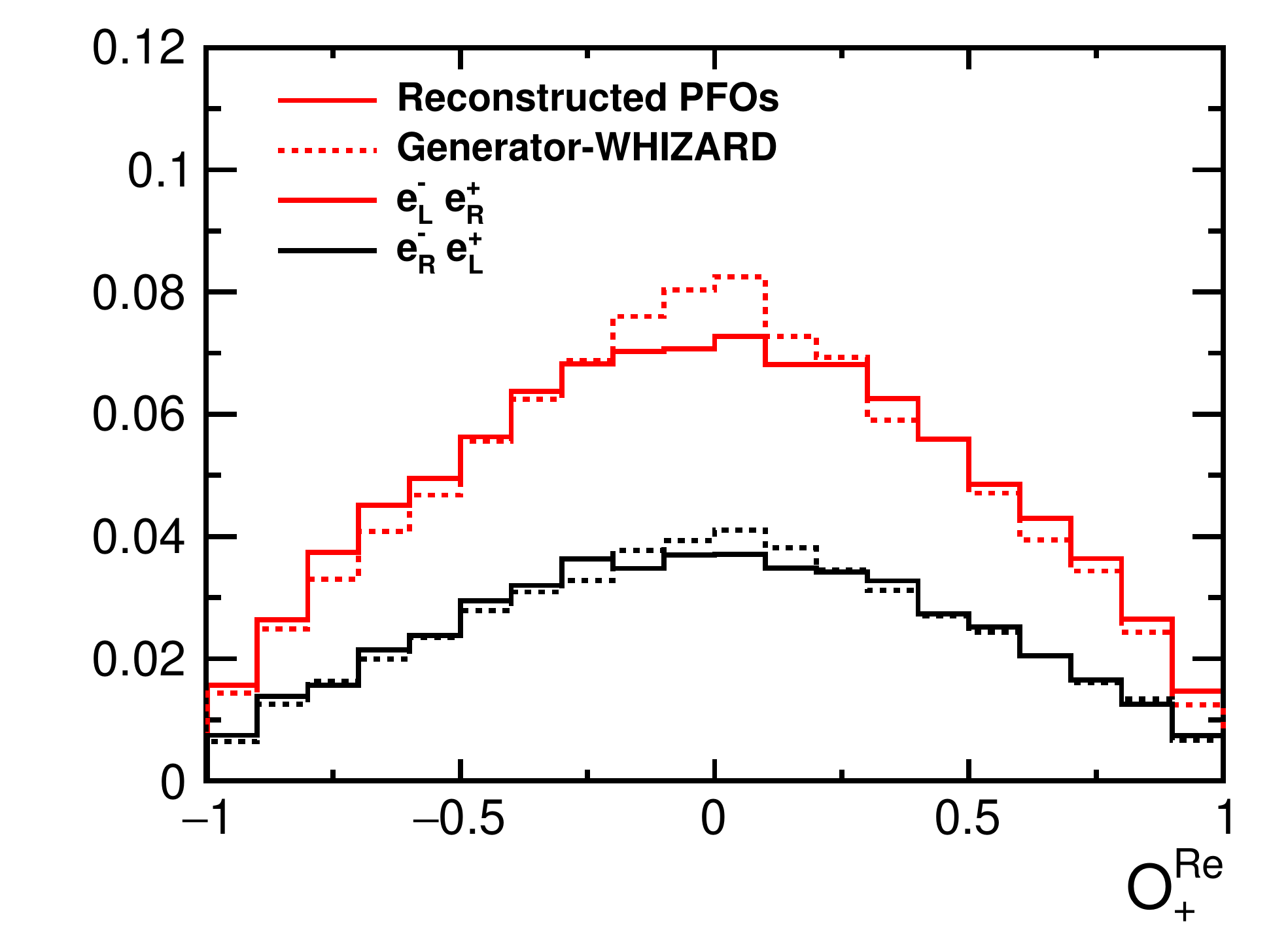}}\hfill 
	\subfloat[\small $\mathcal{O}_{-}^{Re}$]{\includegraphics[width=0.5\textwidth]{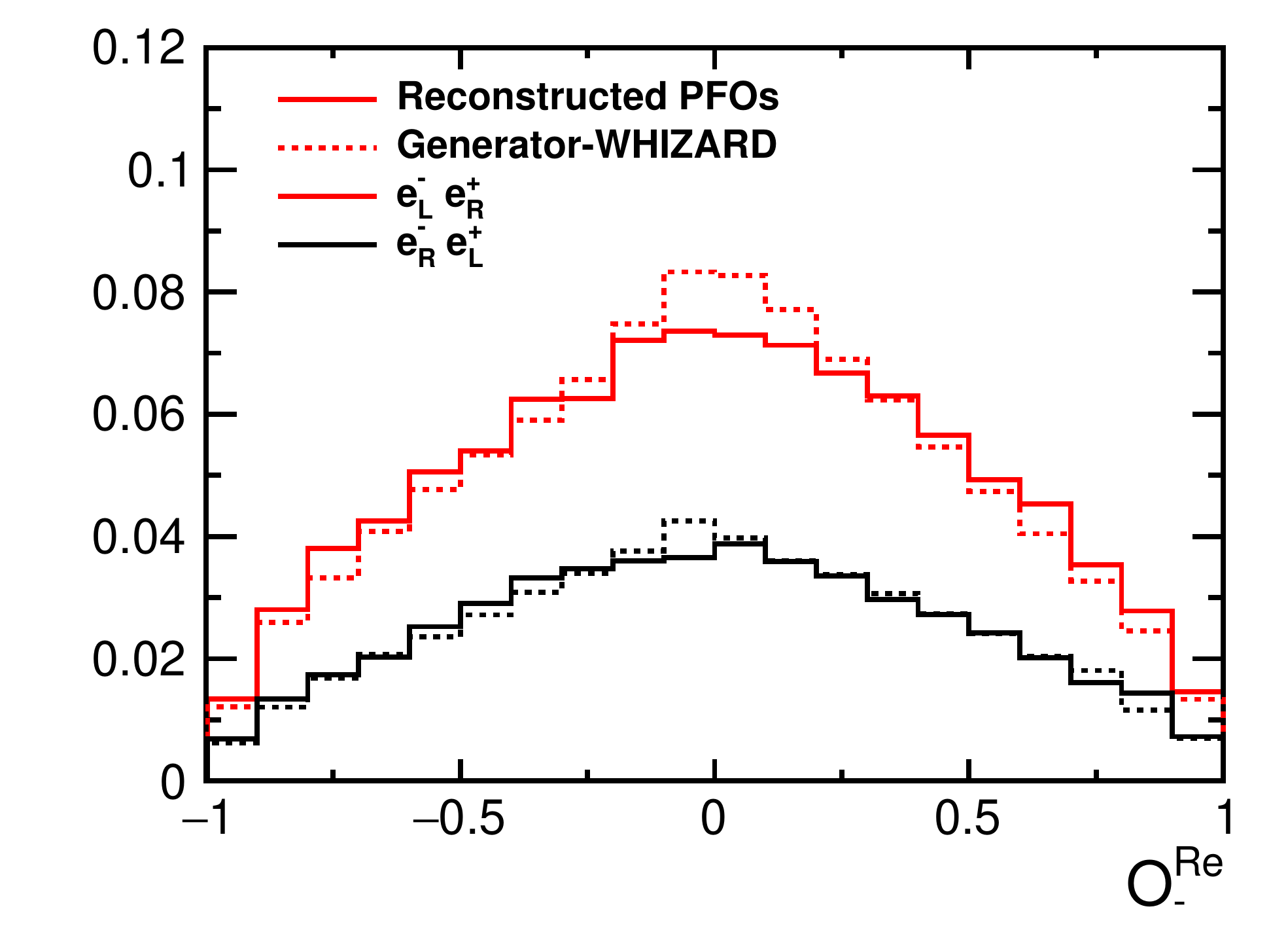}}\\
	\subfloat[\small $\mathcal{O}_{+}^{Im}$]{\includegraphics[width=0.5\textwidth]{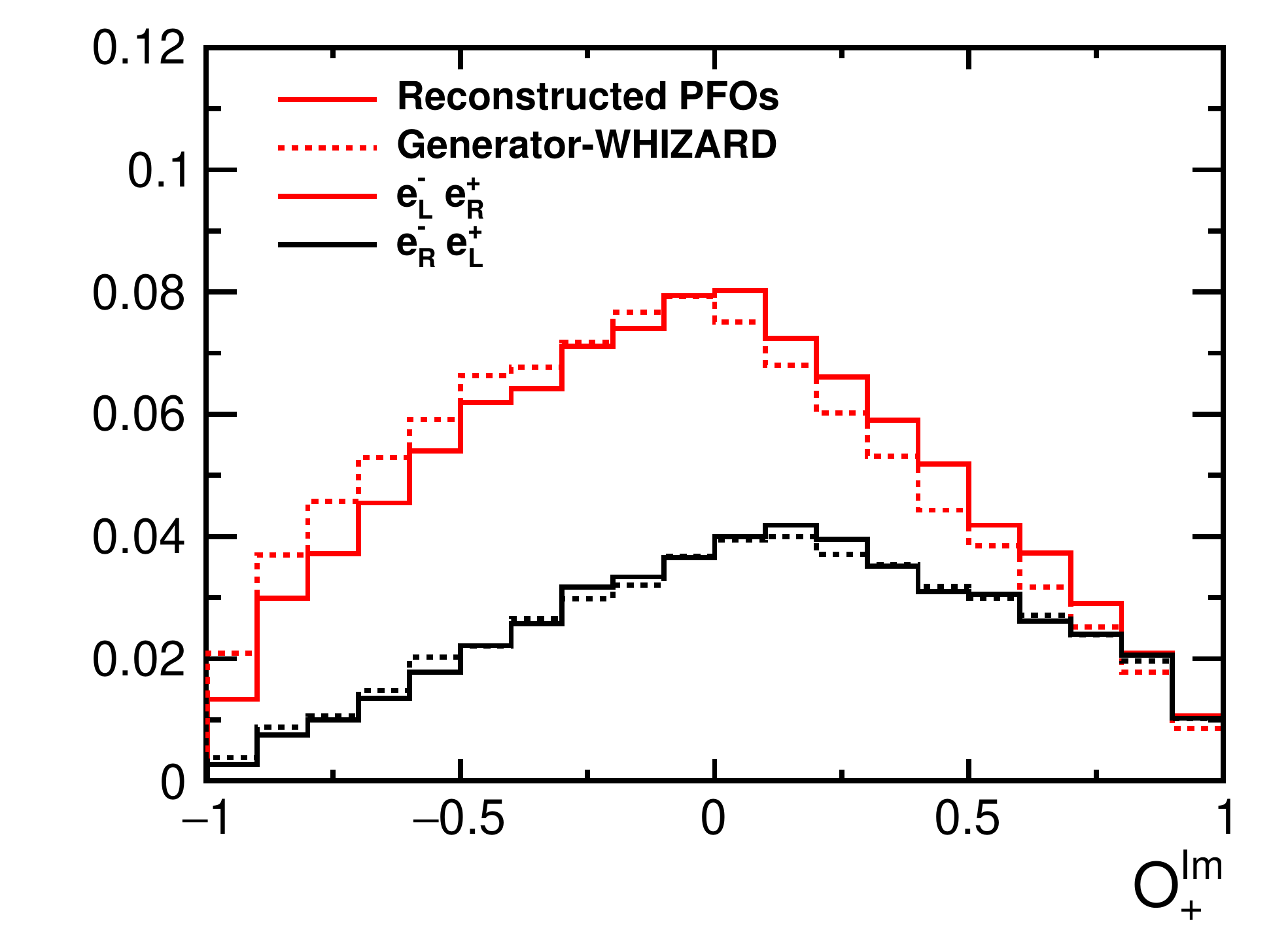}}\hfill 
	\subfloat[\small $\mathcal{O}_{-}^{Im}$]{\includegraphics[width=0.5\textwidth]{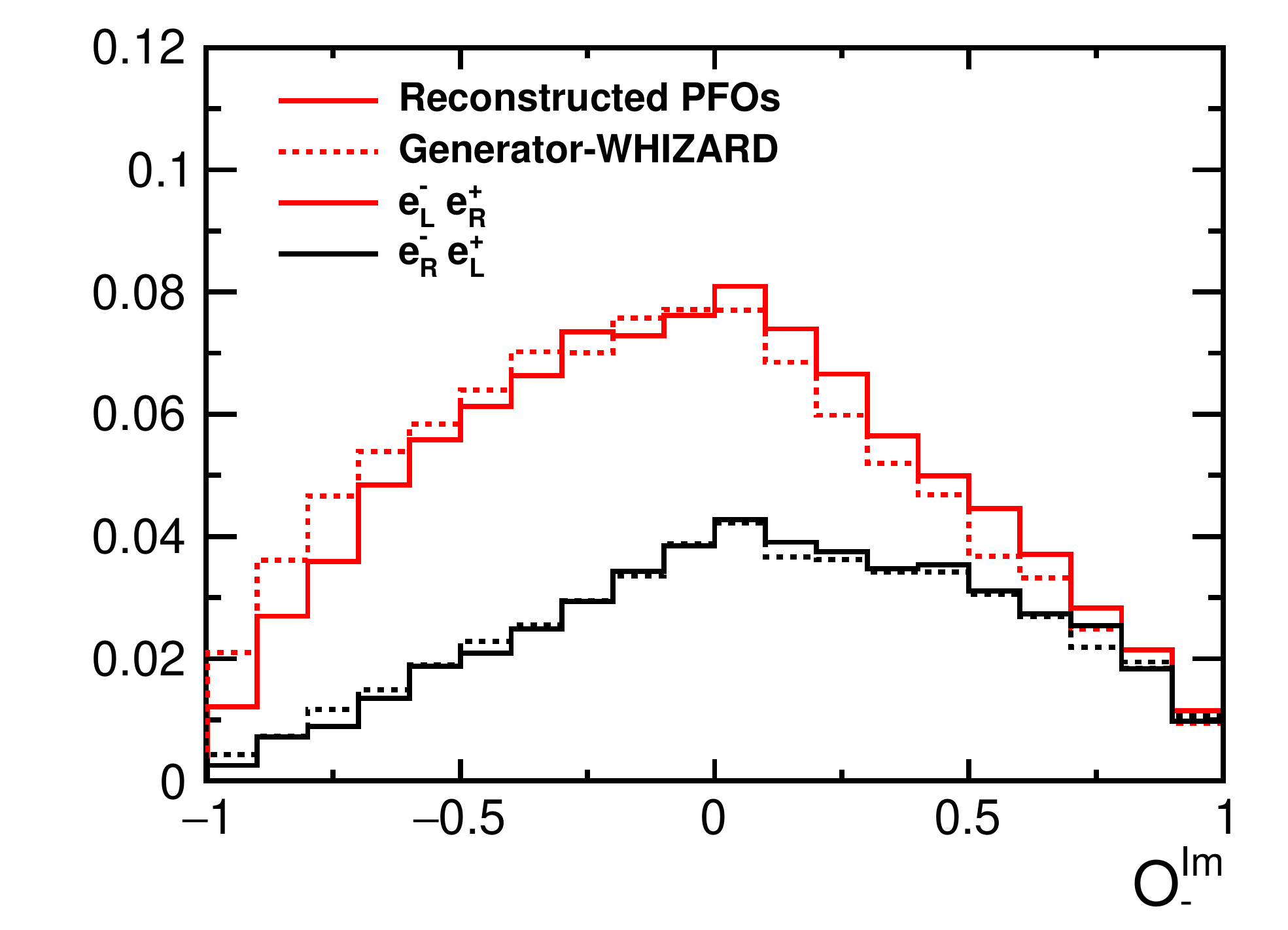}}
	\caption{\small{The CP-odd observables $\mathcal{O}_{\pm}^{Re,Im}$ for the ILC at $\sqrt{s}=$ 500~\gev{}. 
	The four distributions correspond to the reconstructed (solid) and true (dashed) distributions 
	for two beam polarizations. The red histogram ($e^-_L e^+_R$) 
corresponds to -80\% electron polarization and +30\% positron polarization, 
the black histogram ($e^-_R e^+_L$) to +80\% electron polarization and -30\%
positron polarization. The histogram for the
left-handed electron beam is normalized to unit area. The area of the 
histogram for right-handed polarization is scaled so as
to maintain the cross section ratios.
	}
	 }
	\label{fig:cpvobs1}
	}
	\end{figure}

The generated events are processed with the ILD detector
simulation software based on GEANT4~\cite{Agostinelli:2002hh}. 
The ILD detector model is described in the
{\em Detailed Baseline Design} included in the ILC TDR~\cite{Behnke:2013lya}. 
The ILD detector consists of cylindrical {\em barrel} detectors and two
{\em end-caps}. Together these provides nearly hermetic coverage down to
a polar angle of approximately 6 degrees.
For the reconstruction of charged particles ILD relies on
a combination of a solid and gaseous tracking system in a 3.5~Tesla magnetic 
field. Precise silicon pixel and micro-strip detectors occupy the inner radii,
from $r=$ 1.5 cm to $r=$ 33 cm. A large Time Projection Chamber provides 
measurements
out to 1.8~m. The tracker is surrounded by a highly granular calorimeter
designed for particle flow. A highly segmented tungsten 
electromagnetic calorimeter provides up to 30 samples in depth 
with a transverse cell size of 5 $\times$ 5~mm$^2$. 
This is followed by a highly segmented hadronic calorimeter 
with 48 steel absorber layers and 3 $\times$ 3~cm$^2$ read-out tiles. 

The $\gamma \gamma \rightarrow$ {\em hadrons} background corresponding 
to a single bunch crossing is overlaid. 
The data from the different sub-detectors are combined into 
particle-flow objects (PFO) using the Pandora~\cite{Marshall:2012hh}
particle flow algorithm.
Jets are reconstructed using a robust algorithm~\cite{Boronat:2014hva} 
specifically designed for high-energy lepton colliders with non-negligible 
background levels. Particle-flow objects are clustered into exactly
four jets. Heavy-flavour jets are identified using the LCFI 
algorithm~\cite{Bailey:2009ui,Suehara:2015ura}.

The selection and reconstruction of the top-quark candidates 
proceeds as described in Ref.~\cite{Amjad:2015mma}.
The event selection relies primarily on b-tagging and the requirement of
an isolated lepton. The $e^+e^- \rightarrow b\bar{b}l^{\pm}\nu_l q\bar{q}$ 
process includes a small fraction of single-top, that is considered
part of the signal, and less than 1\% of $WWZ$ events. The selection
is based on extensive studies in Refs.~\cite{Amjad:2015mma,Doublet:2012wf}.
The contamination of the signal sample by events due to processes
other than those included in the 
$e^+e^- \rightarrow b\bar{b}l^{\pm}\nu_l q\bar{q}$ sample is less than 5~\% and
is neglected in the following.

The average selection efficiency for signal events 
is approximately 54~\% for the $LR$ sample and 56~\% for the $RL$ polarized
case. The efficiency is over 70~\% for events
with muons and 2~\% lower for events with electrons
or positrons. Events with $\tau$-leptons enter the signal selection
with an efficiency of 20~\%, thanks to $\tau$-decays to electrons
and muons. As expected, no significant difference is observed between 
the selection efficiencies for positively and negatively charged leptons.

The hadronic top candidate is reconstructed
by pairing the two light-quark jets with the b-jet that minimizes
a $\chi^2$ based on the expected $W$-boson and top quark energy and
mass and on the angle between the $W$-boson and the b-jet. 
For $e^-_L e^+_R$ polarization migrations strongly affect the distributions.
A maximum $\chi^2$ is required to retain only well-reconstructed
events. This requirement reduces the overall selection
efficiency to approximately 30\%. This quality cut is not 
applied for $e^-_R e^+_L$ polarization, where migrations have a
small effect.

The reconstructed distributions for the observables $\mathcal{O}_{\pm}^{Re}$ 
and $\mathcal{O}_{\pm}^{Im}$  
are shown in Fig.~\ref{fig:cpvobs1}. In the same figure the true 
distribution is shown, that is, the distribution of
the observable constructed with the lepton and top quark from the Monte 
Carlo record, before any detector effects or
selection cuts are applied.
 
The event selection has a clear impact 
on the  distributions of $\mathcal{O}_{\pm}^{Re}$. A dip in the central 
part of the reconstructed distributions is observed that is due to the limited
acceptance of the experiment in the forward region.
The cuts on lepton energy and isolation have a very small effect. The energy
resolution of the reconstructed hadronic top-quark candidate and 
ambiguities in the assignment of b-jets to W-boson candidates leads
to a slight broadening of the distribution. 
The distributions of $\mathcal{O}_{\pm}^{Im}$ moreover exhibit the expected 
asymmetry due to the beam polarization.

 \begin{table}[h!]
        \begin{center}
  	\caption{Reconstructed values of the CP-odd asymmetries from a Monte Carlo simulation of the ILD detector 
  	response to $\ttbar$ events produced in electron-positron collisions at $\sqrt{s} = $ 500~\gev{}. The quoted uncertainties are due to the limited statisitcs of the simulated samples. }     
	\begin{tabular}{ccc}
		\hline
	polarization  & $e^-_L e^+_R$ ($P_{e^-},P_{e^+} = -0.8, +0.3$) & $e^-_R e^+_L$ ($P_{e^-},P_{e^+} = +0.8, -0.3$)\\ \hline
		$\mathcal{A}^{Re}$& -0.001 $\pm$ 0.003  & -0.009 $\pm$ 0.004\\ 
		$\mathcal{A}^{Im}$&  0.0004 $\pm$ 0.003 & -0.005 $\pm$ 0.004 \\ \hline
	\end{tabular}
	\label{tab:asymcpvILC}
	\end{center}
	\end{table}

The response of the experiment is the same for positively and negatively 
charged leptons and for the hadronic top and anti-top quark 
decay products. Therefore, any distortions 
in the reconstructed distributions are expected to cancel in the asymmetries 
$\mathcal{A}^{Re}$ and $\mathcal{A}^{Im}$. Experimental effects 
generally do not generate spurious asymmetries. The reconstructed 
asymmetries in Table~\ref{tab:asymcpvILC} are found to be compatible with zero
within the statistical uncertainty of $0.003-0.004$.

\section{Full simulation: CLIC at 380~\gev}
\label{sec:CPCLIC}

In this section we study the potential of CLIC operation at $\sqrt s$ = 
380 GeV. The baseline CLIC design allows for up to $\pm$ 80\% longitudinal
electron polarization ($P_{e^-}=\pm 0.8$). Space is reserved in the layout 
for positron polarization as an upgrade option. No positron polarization
is assumed in the following. An integrated 
luminosity of 500~\ifb{} is assumed.

\begin{figure}[h!]
	{\centering
	\subfloat[\small $\mathcal{O}_{+}^{Re}$]{\includegraphics[width=0.5\textwidth]{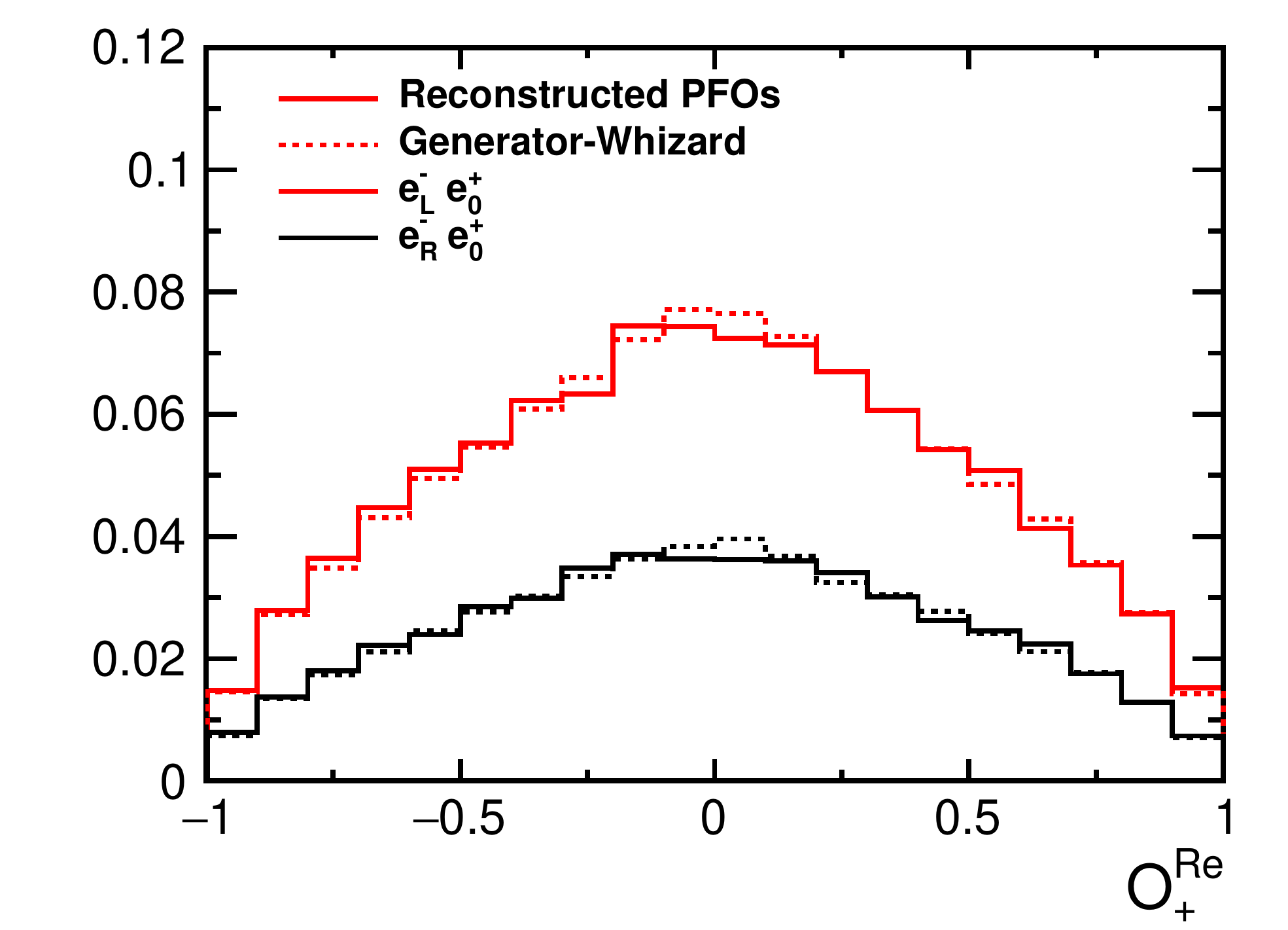}}\hfill 
	\subfloat[\small $\mathcal{O}_{-}^{Re}$]{\includegraphics[width=0.5\textwidth]{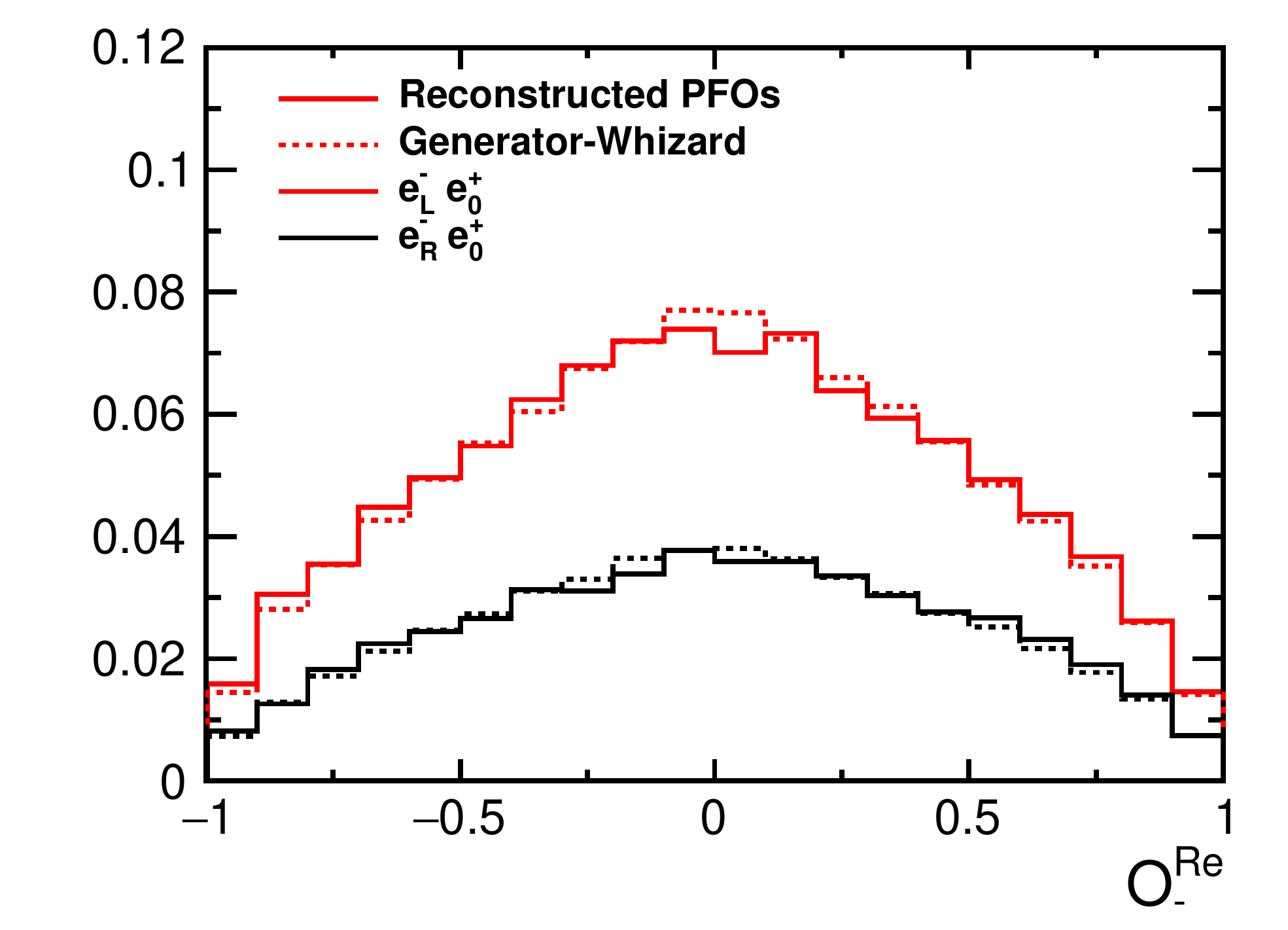}}\\
	\subfloat[\small $\mathcal{O}_{+}^{Im}$]{\includegraphics[width=0.5\textwidth]{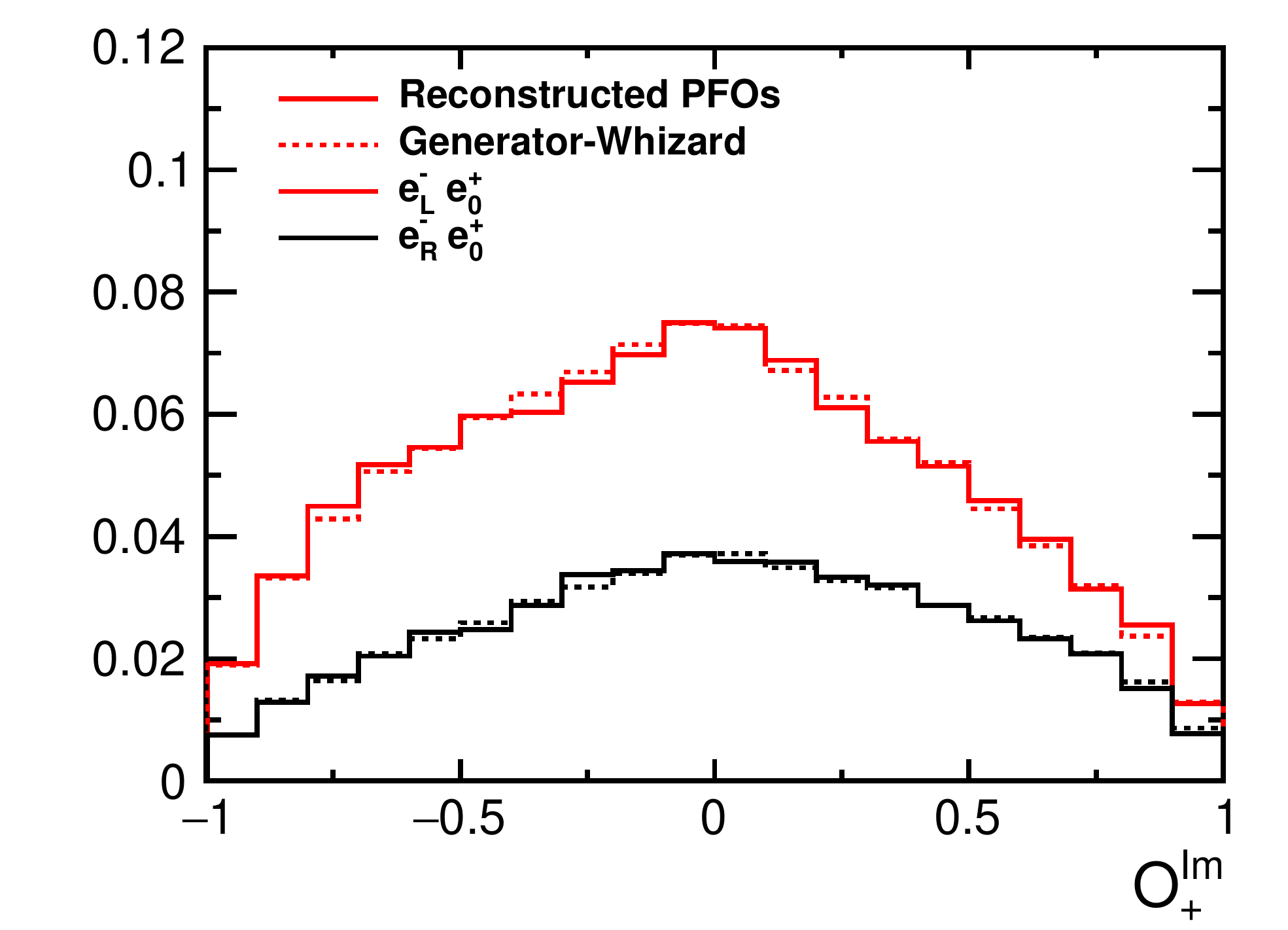}}\hfill 
	\subfloat[\small $\mathcal{O}_{-}^{Im}$]{\includegraphics[width=0.5\textwidth]{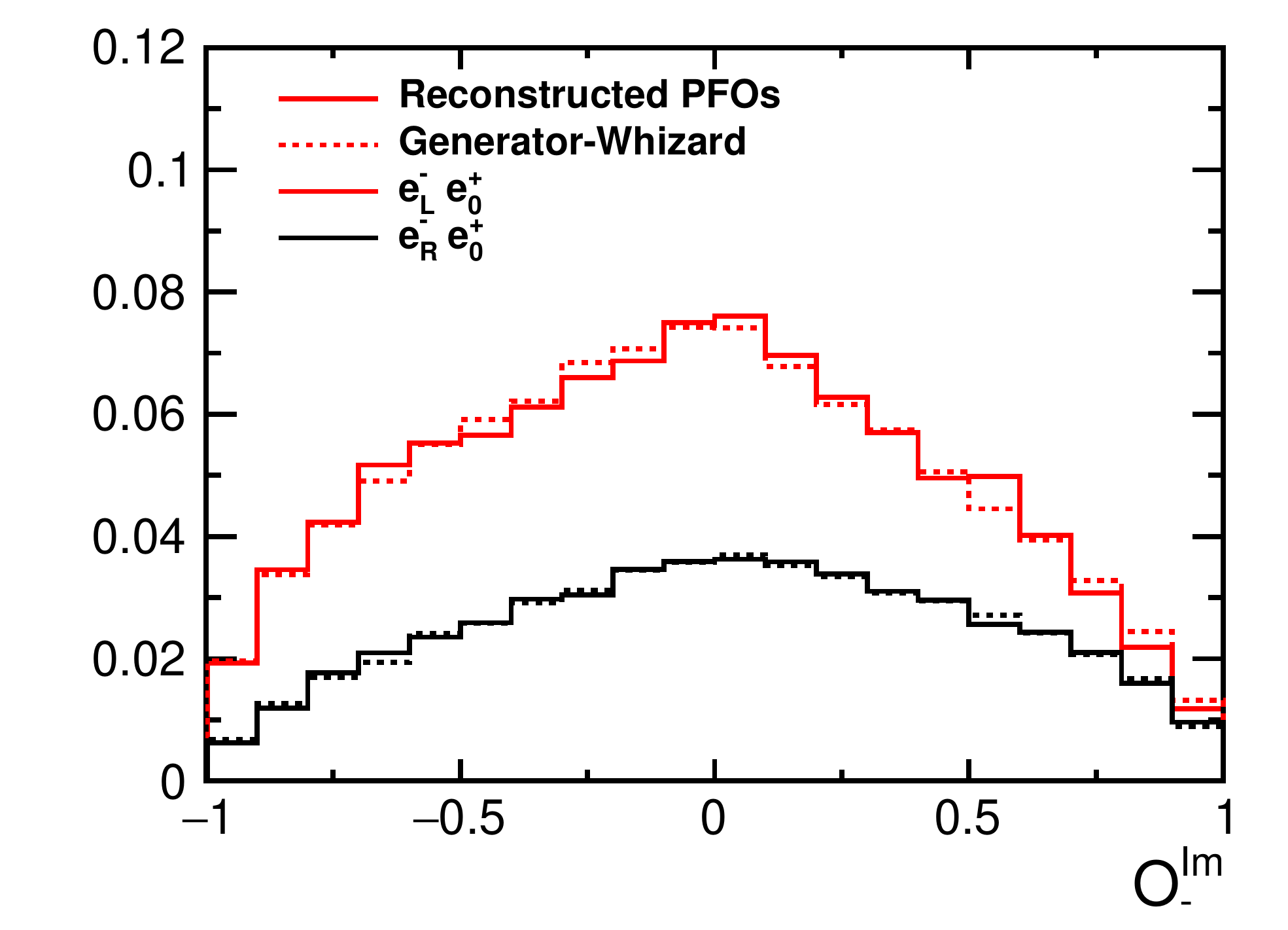}}
	\caption{\small{The CP-odd observables $\mathcal{O}_{\pm}^{Re,Im}$ for CLIC at $\sqrt{s}=$ 380~\gev{}. 
	The four distributions correspond to the reconstructed and true distributions for two beam polarizations. The red histogram ($e^-_L e^+_0$) 
corresponds to -80\% electron polarization, the black histogram ($e^-_R e^+_0$) to +80\% electron polarization. The histogram for the
left-handed electron beam is normalized to unit area. The area of the 
histogram for right-handed polarization is scaled so as
to maintain the cross section ratios. 
	 }
	\label{fig:cpvobs2}
	}
	}
	\end{figure}

Events are generated with WHIZARD 1.95~\cite{Kilian:2007gr}, 
again including all six-fermion processes that produce 
the relevant final state. The effect of ISR and the 
CLIC luminosity spectrum are taken into account. The 
machine parameters correspond to the settings reported 
in the CLIC Conceptual Design Report~\cite{Linssen:2012hp}.

The generated events are processed with a full simulation of 
the CLIC\_ILD detector~\cite{Linssen:2012hp}. The CLIC\_ILD
detector is an adaptation of the ILD detector described in 
Section~\ref{sec:CPILC} to the high-energy environment.
To deal with machine-induced backgrounds the vertex detector is moved out 
to $r=$ 2.5~cm and the time stamping
capabilities of the detector are reinforced. 
The thickness of the calorimeter is enhanced to fully contain energetic
jets: the combination of electromagnetic and hadronic systems 
corresponds to 8.5 interaction lengths.
The electromagnetic calorimeter and barrel 
hadron calorimeter use Tungsten as absorber material.
The end-cap has iron absorber layers. The electromagnetic calorimeter
is read out by 30 sampling layers with finely segmented silicon detectors, 
with a pad size of 5 $\times$ 5 $\mathrm{mm}^2$.
The hadronic calorimeter is read out by 75 layers (60 in the end-cap)
of scintillator material with a cell size of 30 $\times$ 30 $\mathrm{mm}^2$. 

To deal with the background from 
$\gamma \gamma \rightarrow $ {\em hadrons}, particle 
flow objects are selected using a set of timing and energy cuts, 
corresponding to the loose selection of Ref.~\cite{Marshall:2012ry}.

The event selection is identical to that described in Section~\ref{sec:CPILC}.
The b-tagging likelihood cut is reoptimized to achieve a similar signal
efficiency. The overall selection
efficiency is somewhat higher than for the ILC at 500~\gev: 58~\% for the
average over lepton flavours and beam polarizations. The efficiencies
for the two beam polarizations agree within 1~\%. A similar pattern
is observed for the lepton flavours: the efficiency for events with
muons, electrons and $\tau$-leptons are $\sim $ 82\%, $\sim $ 74\%
and $\sim $ 20\%. 

Reconstruction of the $W$-boson and top quark candidates 
proceeds as described in Section~\ref{sec:CPILC}.
At a centre-of-mass energy of 380~\gev{} the 
observables  
are reconstructed quite accurately. The distributions are centered at zero.  
A slight dip is visible at the centre of the reconstructed $\mathcal{O}_{\pm}^{Re}$ distribution due to the limited acceptance in the forward region of the 
experiment.
Other than that the differences between reconstructed and generated distributions are very small.

Again, we find that
the reconstructed asymmetries given in Table~\ref{tab:asymcpvLRCLIC} are 
compatible with zero within the statistical uncertainty. The
entry of $\mathcal{A}^{Re}$ for $P_{e^-} =+0.8$, that is 2 $\sigma$ away from
0, is taken to be a statistical fluctuation. Studies of selection
and reconstruction at parton level with much larger samples fail 
to generate spurious non-zero values for the asymmetry. 

\begin{table}[h!]
        \begin{center}
  	\caption{Reconstructed values of the CP-odd asymmetries from a Monte Carlo simulation of the CLIC\_ILD detector 
  	response to $\ttbar$ events produced in electron-positron collisions at $\sqrt{s} = $ 380~\gev{}. The quoted uncertainties are due to the limited statisitcs of the simulated samples.}     
	\begin{tabular}{ccc}
		\hline
	polarization  & $e^-_L$ ($P_{e^-} =-0.8$) & $e^-_R$ ($P_{e^-} =+0.8$)\\ \hline
		$\mathcal{A}^{Re}$& -0.001 $\pm$ 0.003  & 0.009 $\pm$ 0.004\\ 
		$\mathcal{A}^{Im}$& 0.0003 $\pm$ 0.003 & -0.002 $\pm$ 0.003 \\ \hline
	\end{tabular}
	\label{tab:asymcpvLRCLIC}
	\end{center}
	\end{table}

\section{Parton level study for high-energy operation}
\label{sec:CPCLIC3000}

In this Section we study the potential of the high-energy stages 
of the CLIC programme that could reach 3~\tev. The instantaneous
luminosity scales approximately proportional to the centre-of-mass
energy and one may expect an integrated luminosity of several~\iab.

The decay of boosted top quarks produces a topology~\cite{Abdesselam:2010pt} 
that is very different from that of $t{\bar t}$ events close to 
the production threshold. Therefore, the reconstruction of the 1-3~\tev{}
collisions must be performed with an algorithm specifically developed for 
high energy, where the collimated decay products of the hadronic top quark
are captured in a single large-$R$ jet (i.e. a jet reconstructed 
with a radius parameter $R$ greater than 1). In this reconstruction scheme  
the combinatoric problem of pairing $W$-boson and $b$-tagged jets is
entirely avoided.

The $\gamma \gamma \rightarrow $ {\em hadrons} background 
in multi-\tev{} collisions is more severe than at low energy.
The reconstruction of boosted top quarks at CLIC was studied in a 
detailed simulation, including realistic background levels in 
Ref.~\cite{Boronat:2016tgd}. With tight pre-selection cuts on 
the particle-flow objects and the robust algorithm of 
Ref.~\cite{Boronat:2014hva} the top-quark energy can be reconstructed
with a resolution of 8\%. Also the jet mass and other substructure observables
can be reconstructed precisely, with much better resolution than at the LHC.
As background processes have cross sections that are similar to that of 
top-quark production, it seems safe to assume that $\ttbar$ events with
centre-of-mass energies of 1-3~\tev{} can be efficiently selected and
distinguished from background processes.

\begin{figure}[h!]
	{\centering
	\subfloat[\small $\mathcal{O}_{+}^{Re}$]{\includegraphics[width=0.49\textwidth]{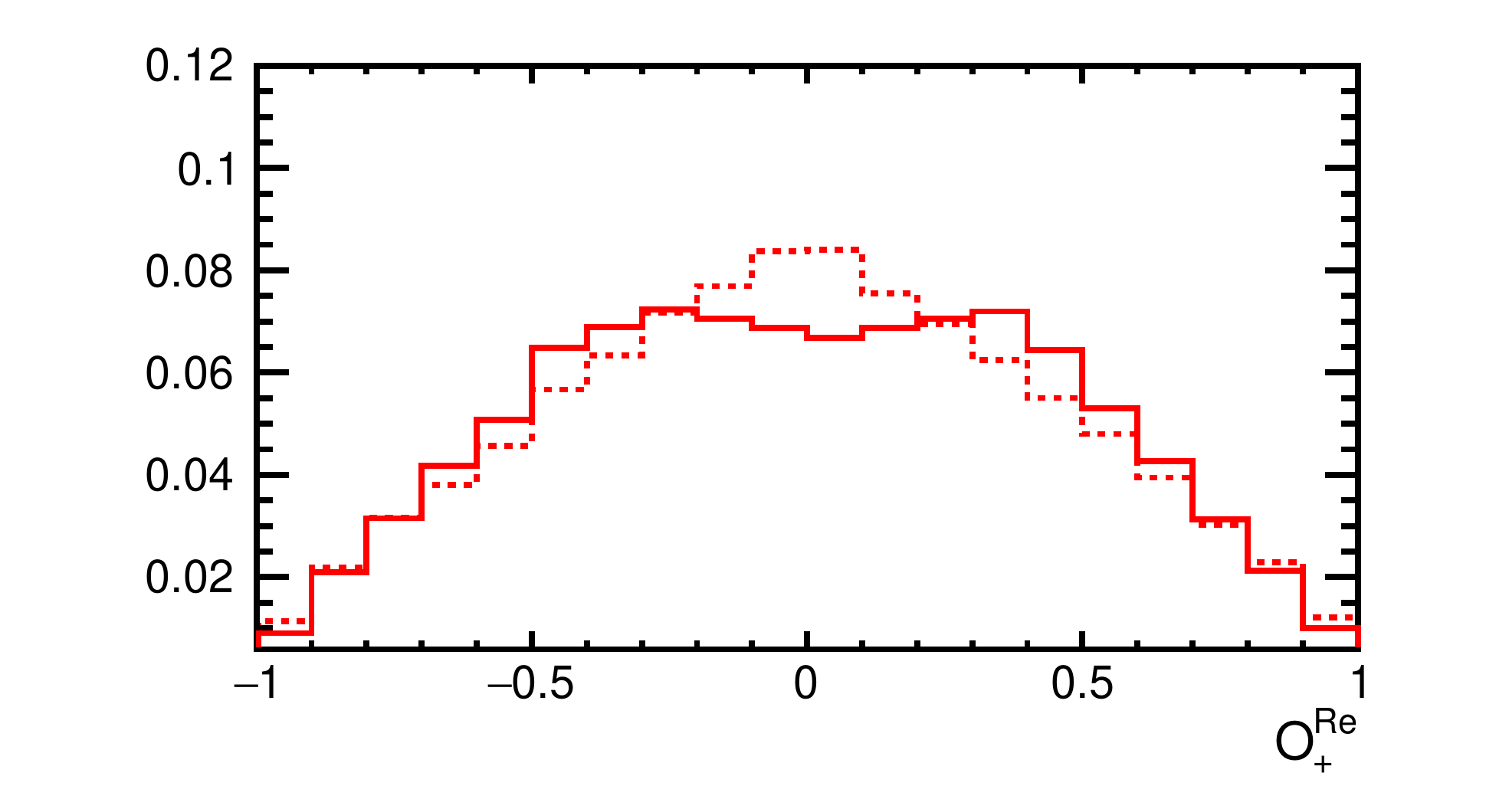}}\hfill 
	\subfloat[\small $\mathcal{O}_{-}^{Re}$]{\includegraphics[width=0.49\textwidth]{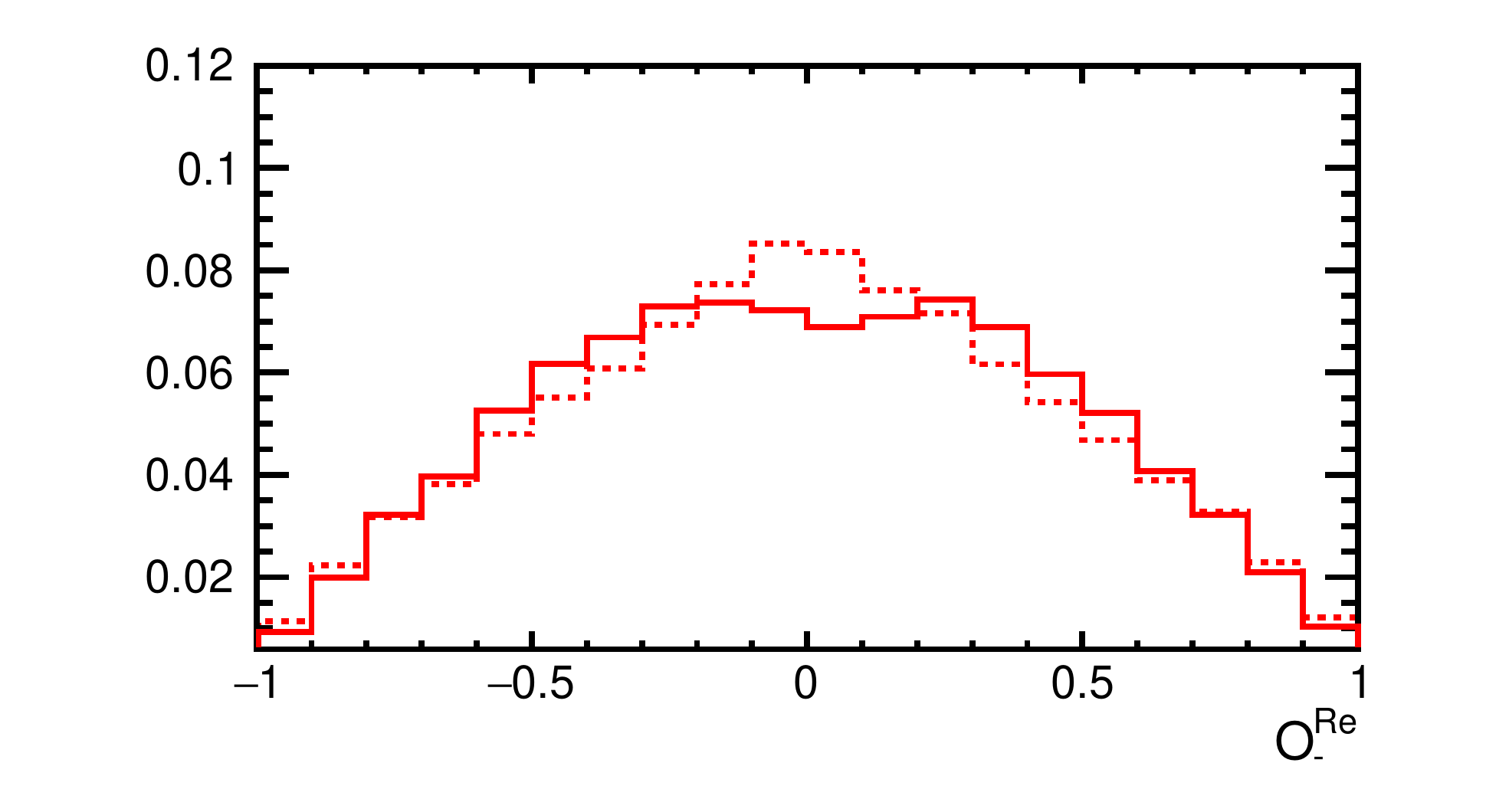}}\\
        \subfloat[\small $\mathcal{O}_{+}^{Re}$]{\includegraphics[width=0.49\textwidth]{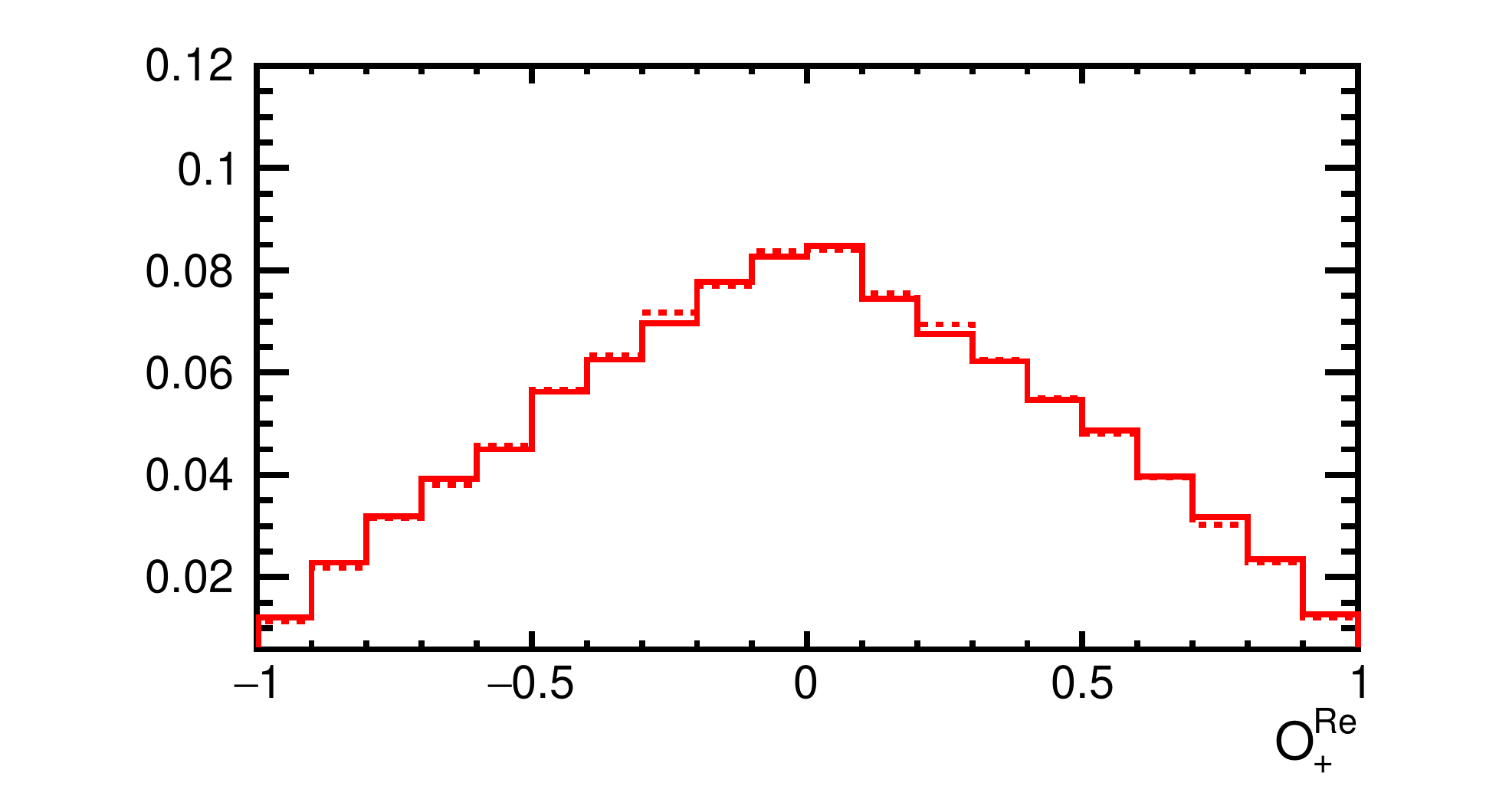}}\hfill 
	\subfloat[\small $\mathcal{O}_{-}^{Re}$]{\includegraphics[width=0.49\textwidth]{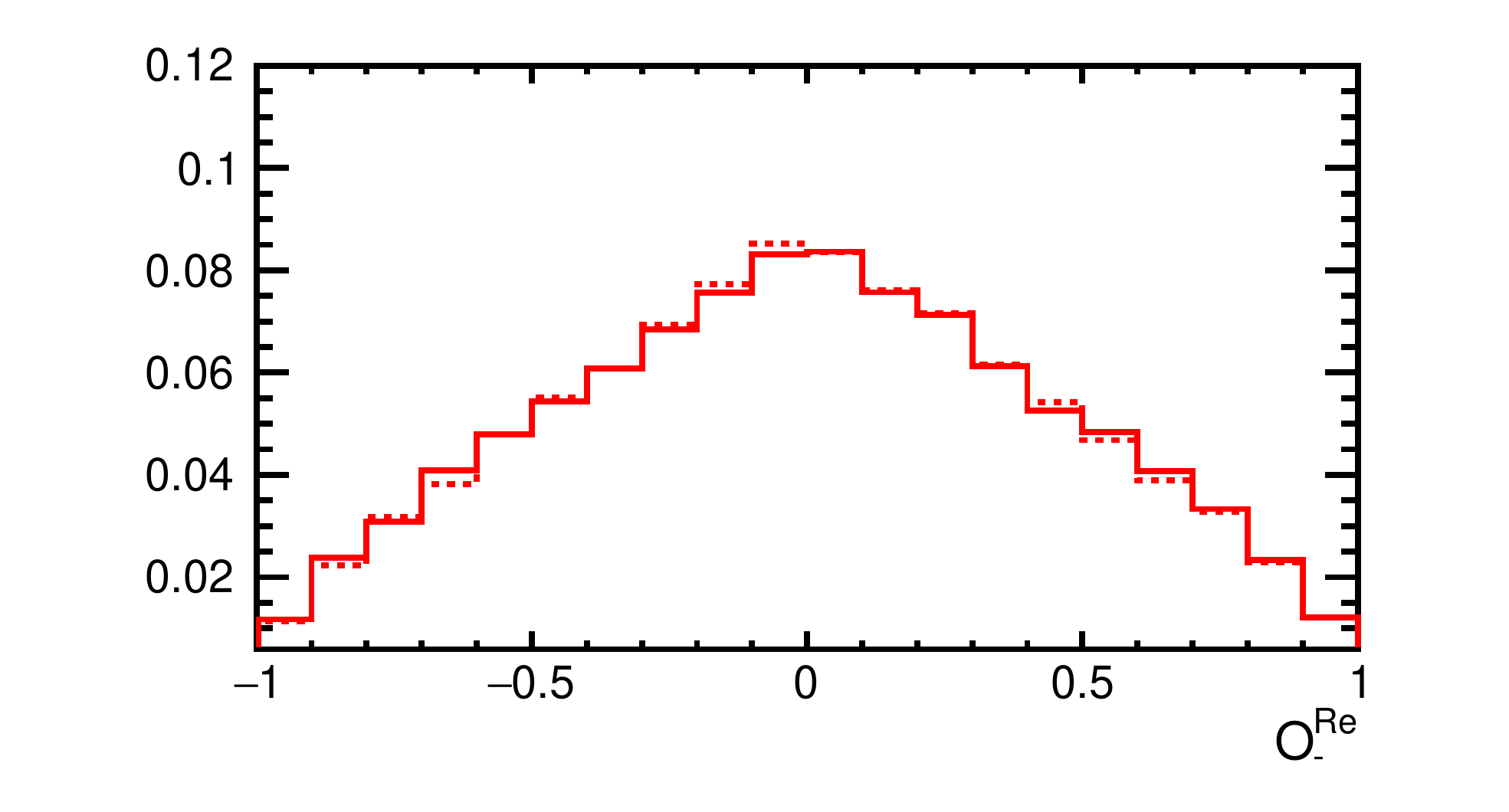}}\\
	\subfloat[\small $\mathcal{O}_{+}^{Im}$]{\includegraphics[width=0.49\textwidth]{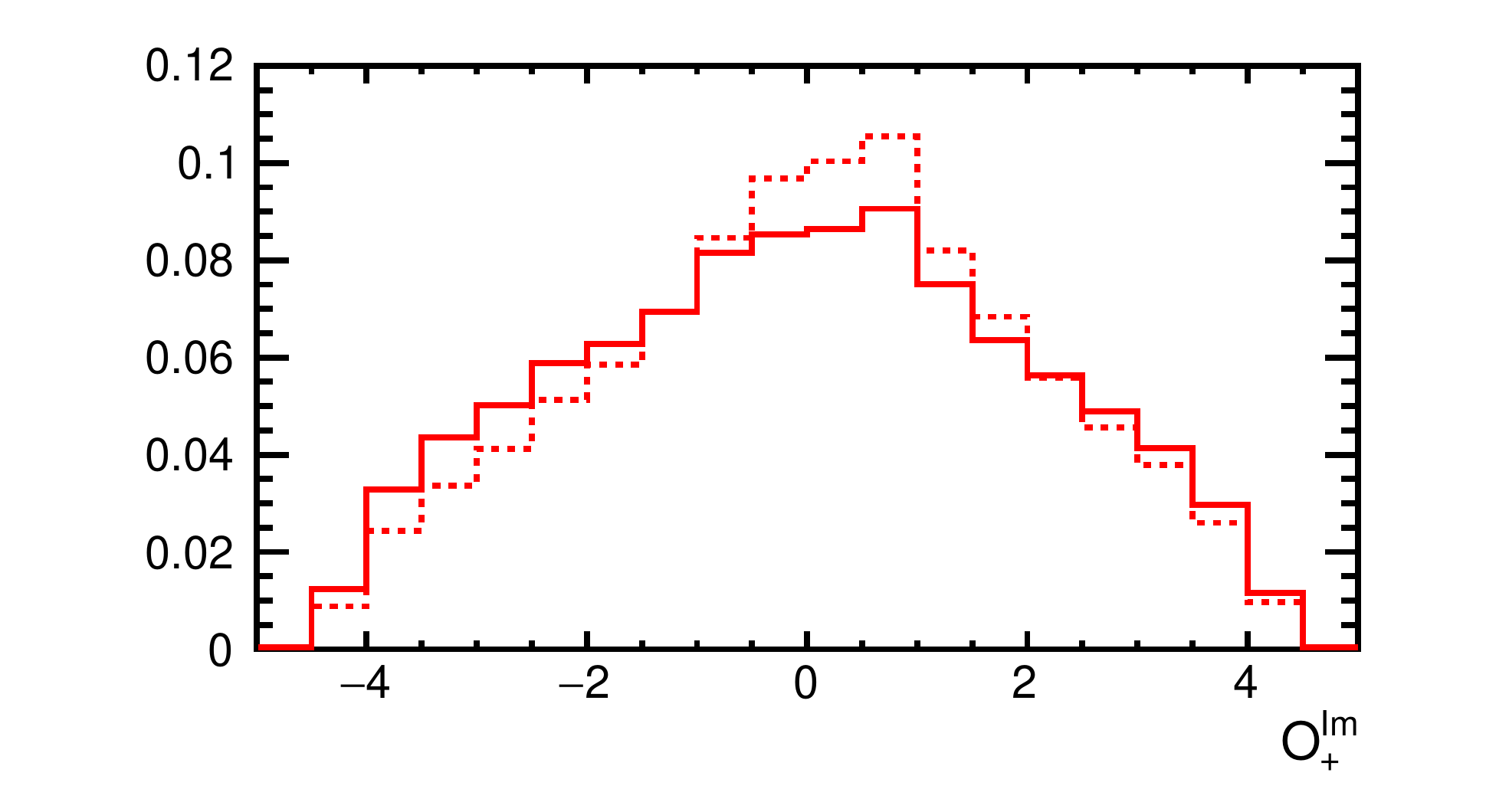}}\hfill 
	\subfloat[\small $\mathcal{O}_{-}^{Im}$]{\includegraphics[width=0.49\textwidth]{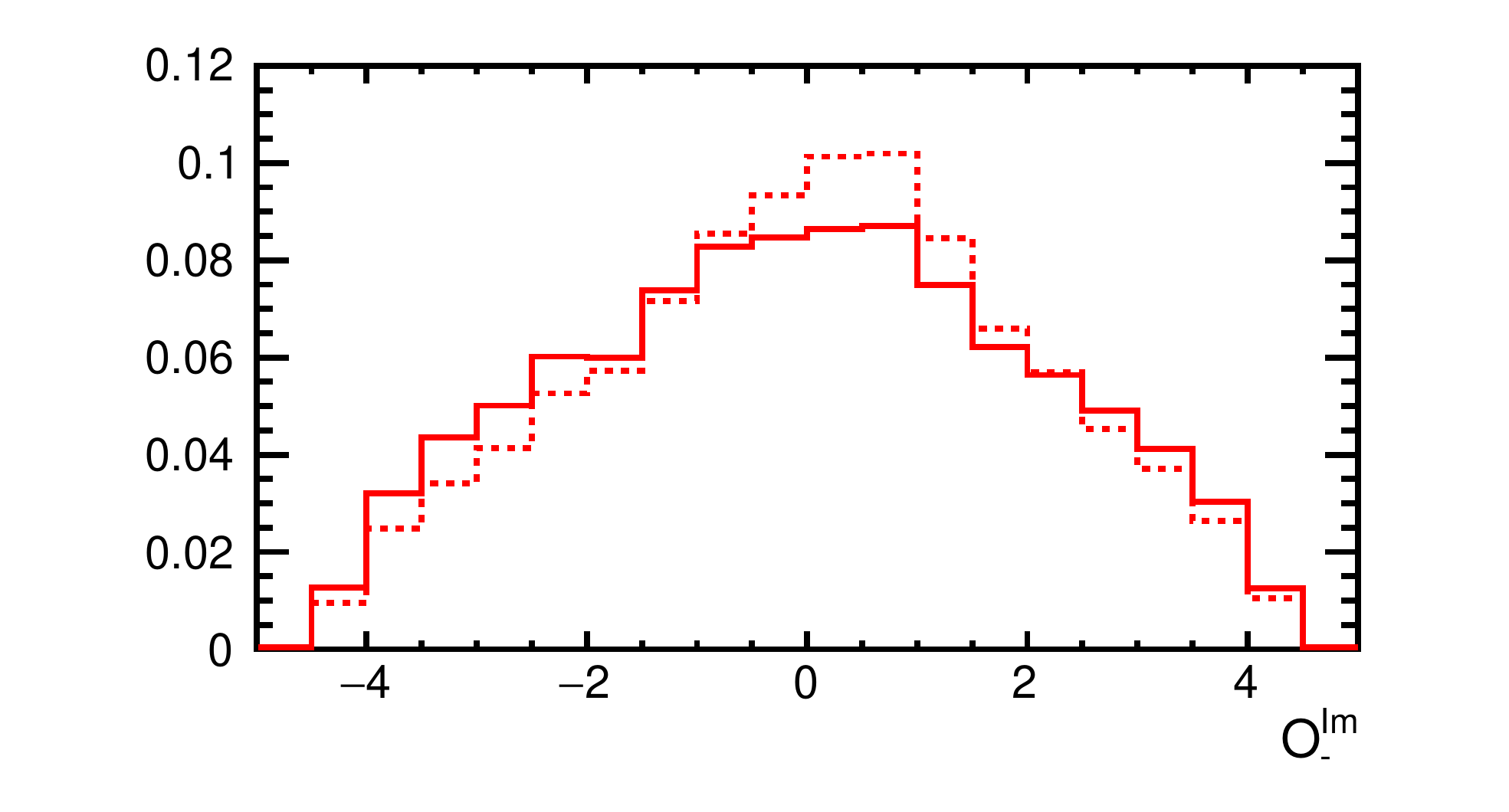}} \\
	\subfloat[\small $\mathcal{O}_{+}^{Im}$]{\includegraphics[width=0.49\textwidth]{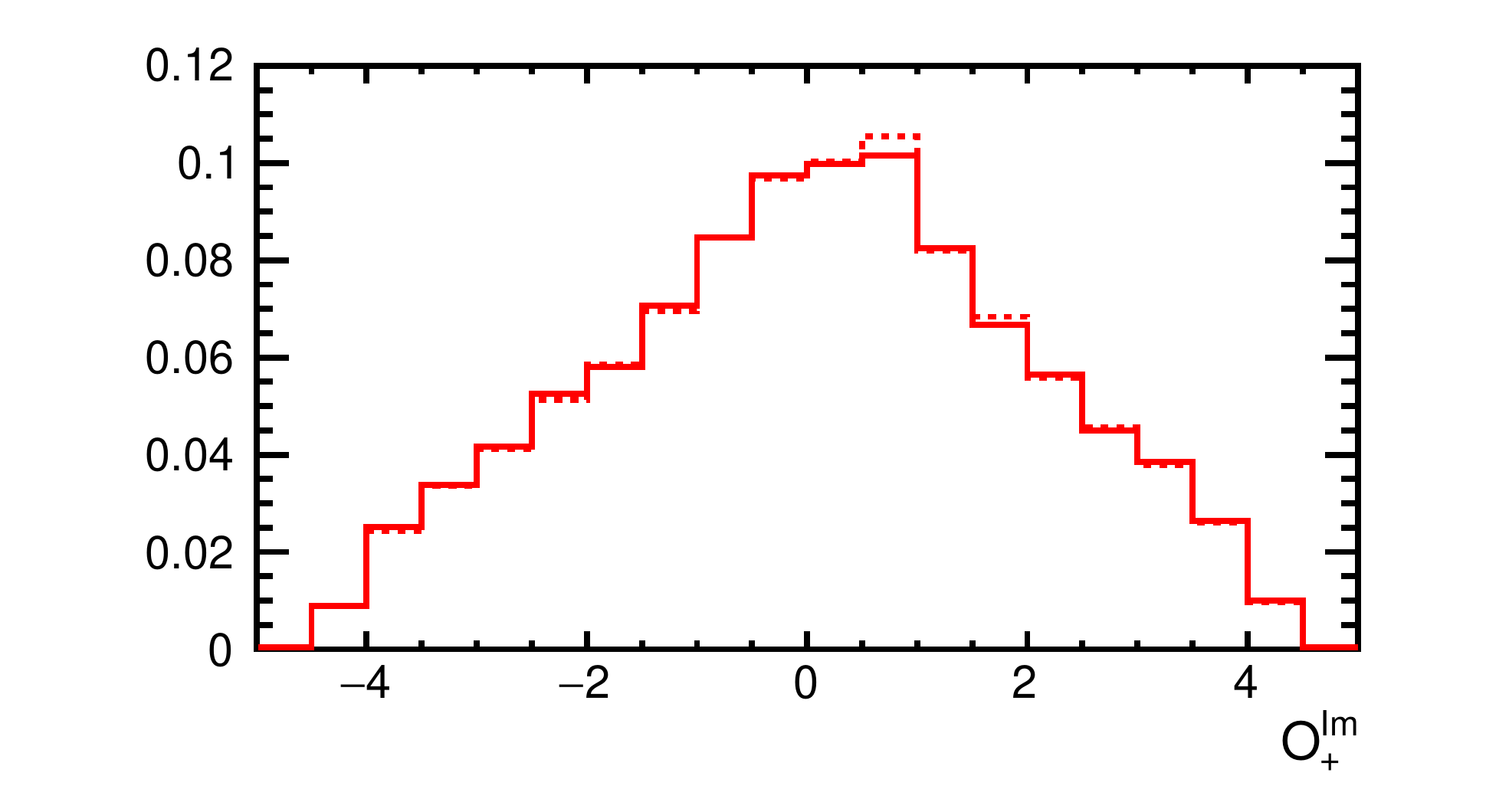}}\hfill 
	\subfloat[\small $\mathcal{O}_{-}^{Im}$]{\includegraphics[width=0.49\textwidth]{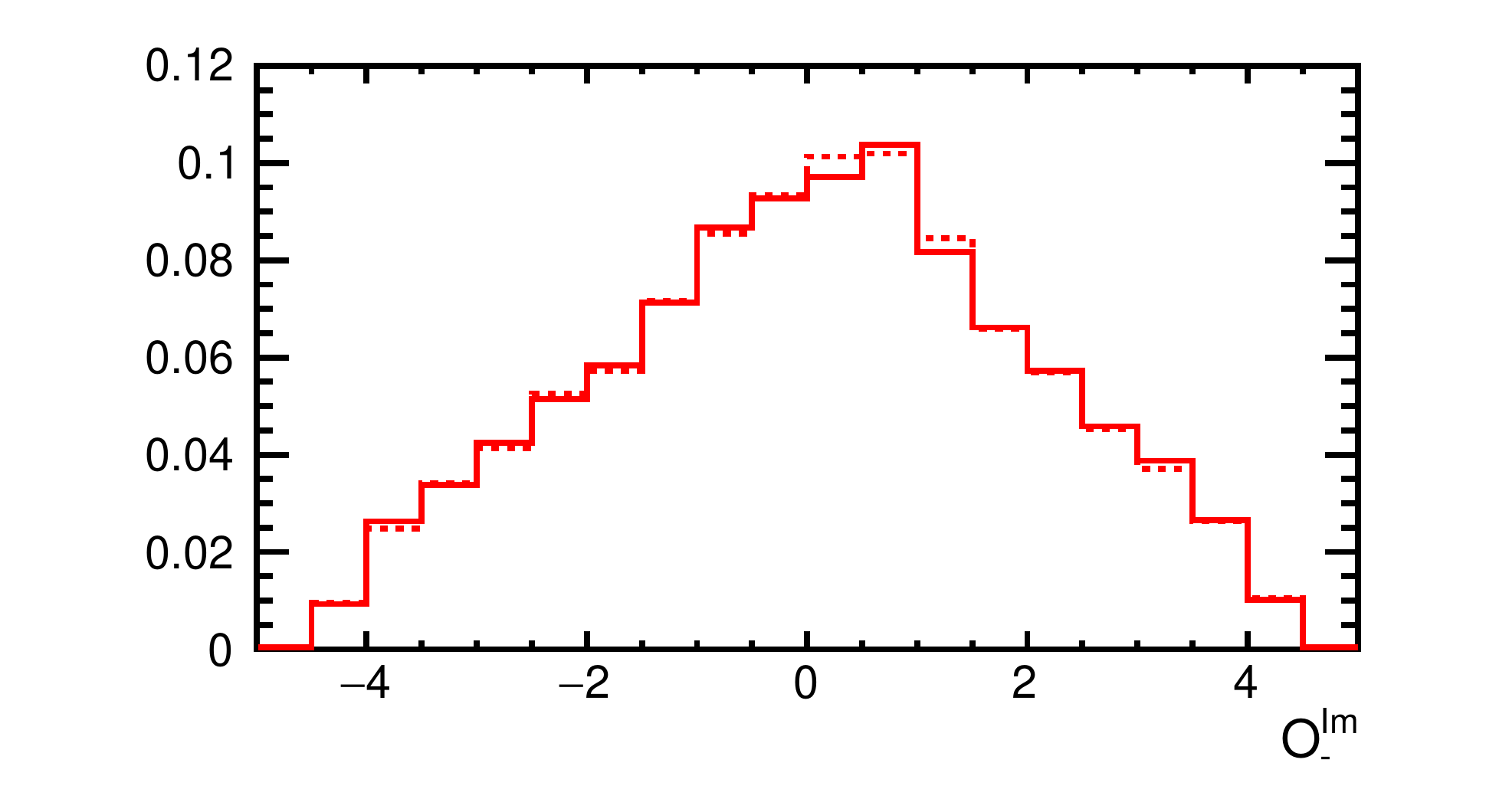}} \\
	\caption{\small{The CP-odd observables $\mathcal{O}_{\pm}^{Re,Im}$ for CLIC at $\sqrt{s}=$ 3~\tev{}. The two 
distributions correspond to the reconstructed (solid) and true (dashed) distributions for $e^-_L e^+_0$ polarization. The results
for $e^-_R e^+_0$ polarization are similar. Panels (a) and (b) represent the effect of the polar angle selection on $\mathcal{O}_{\pm}^{Re}$. In panels (c) and (d) the impact of the angular and energy resolution on $\mathcal{O}_{\pm}^{Re}$ are shown.
 Panels (c) and (d) represent the effect of the selection on $\mathcal{O}_{\pm}^{Im}$, (g) and (h) the impact of the angular and energy resolution. 
All histograms are normalized to unit area. The $\mathcal{O}_{\pm}^{Im}$ distribution is truncated to the interval [-5,5].
	 }
	\label{fig:cpvobs3}
	}
	}
	\end{figure}

An evaluation based on a detailed simulation of the experimental response 
for the optimal observables is not yet available. We identify the most important
effects using a parton-level simulation. A representative selection is applied to
parton-level $\epem \rightarrow \ttbar \rightarrow b\bar{b} q \bar{q}' l \nu$ 
events generated with MG5\_aMC@NLO~\cite{Alwall:2014hca}. 
The detector resolution is
implemented by smearing of the parton four-vectors. 

The limited acceptance in the
forward region shapes the distributions significantly. 
For partons emitted at shallow angle, part of the jet energy
flow disappears down the beam pipe. We mimic
this effect by requiring that all partons have $|\cos \theta| <$  0.98 
(the detector coverage extends to well beyond $|\cos \theta| =$ 0.99 ; 
some margin is added as jets have a finite size).
In Figure~\ref{fig:cpvobs3} the distribution for selected events
is compared to the full distribution. The effect is more pronounced
and more localized than in the low-energy analysis.  

We furthermore apply a smearing to mimic the resolution for the hadronic 
top quark
candidate. The reconstructed top-quark four-vector is used to boost the lepton 
to the top-quark system. The finite energy resolution and angular resolution may
lead to distortions of the reconstructed distribution. The effects of a 10\%
energy resolution and 0.02 radian angular resolution, twice the size of
the resolution found in the study of Ref.~\cite{Boronat:2016tgd},
are indicated in Figure~\ref{fig:cpvobs3}. The reconstruction has a much
less severe impact than in the low-energy analysis.

As for the low-energy analysis, these experimental effects are identical for
positively and negatively charged leptons and for quarks and anti-quarks.
We therefore expect that experimental effects do not create spurious 
asymmetries. Rough, but conservative, limits on systematic 
effects are presented in the next section.

A more detailed study on a detailed detector simulation is required
for a quantitative study of the high-energy performance. 
In the following we estimate the potential of high-energy operation, 
assuming an acceptance of 40\% for lepton+jets events.

\clearpage

\section{Systematic uncertainties}
\label{sec:systematics}

Before we discuss the prospects of linear colliders to extract the real 
and imaginary parts of the  form factors $F_{2A}^{\gamma,Z}$, a number of
potential sources of systematic uncertainties are briefly discussed. 

The polarization of the electron and positron beams is the key machine 
parameter in the extraction of the form factors. A combination of polarimeters
and in-situ measurements allows for a precise determination of $P_{e^-}$ and 
$P_{e^+}$. The detailed study of the ILC case in Ref.~\cite{Beckmann:2014mka} 
envisages a determination to the 10$^{-3}$ level. The study of (single) $W$-boson
production is expected to provide per-mille level precision at high
energy. This precision is well beyond what is
needed to avoid significant uncertainties in the form factor extraction.
The uncertainties of other machine parameters, such as the integrated luminosity
or the centre-of-mass energy, have a negligible effect on the result.

The analysis is found to be quite robust against the effects of event
selection and reconstruction of the $t\bar{t}$ system. The
limited acceptance and efficiency do lead to significant distortions
of the distributions of $\mathcal{O}_{\pm}^{Re}$ and $\mathcal{O}_{\pm}^{Im}$.
Also, the impact of migrations is clearly visible in each of the 
distributions. However, these effects cancel in the asymmetry.
Therefore, none of these effects generate a non-zero asymmetry when the 
true value is 0. This type of uncertainty is referred to as {\em bias}.
The full-simulation study shows that a spurious non-zero
result due to systematic effects is expected to be smaller than 0.005.

For arbitrary values of the true asymmetry the analysis of the systematics
is a bit more involved. We must also consider the possibility that the
selection and reconstruction of the events 
lead to a non-linearity in the response to non-zero CP 
asymmetries $\mathcal{A}^{Re}$ and $\mathcal{A}^{Im}$. 
These effects are labelled as {\em non-linearity} in the following. They 
are evaluated in a parton-level study
using events generated with non-zero WDF and EDF. 
Distributions and asymmetries with non-zero values of the top-quark 
EDF and WDF are generated using a MadGraph~\cite{Alwall:2014hca} UFO model 
developed in Ref.~\cite{Gupta:2009wu}.
The most important cuts in the analysis, namely
on the charged lepton energy, its isolation, and the polar angle of 
final-state quarks are applied to the six-fermion final state. The finite
resolution in the reconstruction of the hadronic top-quark candidate 
is implemented by smearing the top-quark three-momentum vector. The
migrations due to ambiguities in pairing b-jets and $W$-bosons at low energy
are simulated by implementing the incorrect pairing for 15\% of events. 
The selection tends to enhance the
reconstructed asymmetry. This effect is particularly pronounced
at very high centre-of-mass energy, where it can reach up to 
10\% of the true asymmetry (for $\sqrt{s}=$ 3~\tev). 
Migrations and resolution effects dilute the asymmetry,
yielding reconstructed values that are reduced by 5-15\%.
For centre-of-mass energies of 380~\gev{} or 500~\gev{} migrations
are the most important systematic effect. At higher energy the
resolution is the dominant effect.

Theory uncertainties are estimated as follows. 
Radiative corrections to $\ttbar$ production in $e^+e^-$ collision are known
 to high precision. The next-to-leading order (NLO) QCD corrections have 
been known for a long time~\cite{Jersak:1981sp}. The NLO electroweak 
corrections were determined in 
Refs.\cite{Beenakker:1991ca,Fleischer:2003kk,Hahn:2003ab}.
Off-shell  $\ttbar$ production and decay including non-resonant and 
interference contributions at NLO QCD were investigated in 
Ref.\cite{Nejad:2016bci}.
The NNLO QCD corrections to $\ttbar$ production, including differential 
distributions, were calculated in \cite{Gao:2014eea,Chen:2016zbz}.
Although not done in this work, the coefficients of 
${\rm Re} F_{2A}^{\gamma,Z}$ and ${\rm Im} F_{2A}^{\gamma,Z}$ in the asymmetries of
Equations~\eqref{eq:ARe} and \eqref{eq:AIm} can be computed at NLO in 
the SM couplings. We can then estimate the theory uncertainties of these 
coefficients as follows. The uncertainty of the  $\ttbar$ cross section 
associated with  renormalization scale variations in the range 
$\sqrt{s}/2\leq \mu \leq 2\sqrt{s}$ is at NLO (NNLO) QCD about 
$2\%$ $(1\%)$ at $\sqrt{s}=380$ GeV and $\sim 0.9\%$ $(0.2\%)$ at 
$\sqrt{s}=$ 500~GeV \cite{Chen:2016zbz}.
Assuming that the NLO SM corrections to the 
squared matrix element including the EDF and WDF to $\ttbar$ production 
and decay are known, we take these NLO QCD
 values as theory uncertainties. They are labeled ``theory (non-linearity)'' 
in Table~\ref{tab:systematics}. We believe that these uncertainty estimates 
are not unrealistic because the uncertainties of these coefficients are, 
in fact, associated with the expectation values 
$\langle\mathcal{O}_{\pm}^{Re}\rangle$, $\langle\mathcal{O}_{\pm}^{Im}\rangle$, 
which are ratios that are usually expanded in powers of the SM couplings. 
QCD scale uncertainties of expanded ratios are in general smaller 
than the scale uncertainty of the cross section. 
An example is the top-quark forward-backward asymmetry $A_{\rm FB}^t$ 
which is known to NNLO QCD accuracy \cite{Gao:2014eea,Chen:2016zbz}. 
The scale uncertainty  of the expanded $A_{\rm FB}^t$ is below $0.5\%$
at these c.m. energies \cite{Chen:2016zbz}.
       
The numbers in the row ``theory (bias)'' in Table~\ref{tab:systematics} 
are a very conservative estimate of CP-violating SM contributions 
induced by higher-order $W$-boson exchange to $e^+e^-\to t{\bar t}$.  
At one loop in the electroweak couplings there are no CP-violating SM 
contributions to this flavour-diagonal reaction. Beyond one loop the 
CP-violating SM contributions to the asymmetries of Equations~\eqref{eq:ARe}, 
\eqref{eq:AIm}  are smaller than $[g_W^2/(16\pi^2)]^2 {\rm Im} J$, 
where $g_W=e/\sin\theta_W$ and ${\rm Im} J$ is the imaginary part 
of a product of four quark mixing-matrix elements, which is invariant 
under phase-changes of the quark fields. 
Its value is $|{\rm Im} J| \sim 2\times 10^{-5}$.

\begin{table}
  \begin{center}
   \caption{The main systematic uncertainties on the asymmetry 
   $\mathcal{A}^{Re}$ for left-handed polarized electron beam 
   (and right-handed positron beam in the case of 500~\gev{} operation).
   Entries labelled {\em bias} represent estimates of upper bounds on 
   systematic effects that yield a spurious non-zero result 
   in the Standard Model. Entries labelled {\em non-linearity}
   represent systematic uncertainties that affect the proportionality
   of the response to non-zero values of the asymmetry (induced by
   physics beyond the Standard Model). Positive signs indicate effects
   that enhance the observed asymmetry. Negative signs corresponds
   to effects that dilute the asymmetry. 
} 
    \begin{tabular}{lccc}
      \hline
      source & 380~\gev{} & 500~\gev{} & 3~\tev{} \\ \hline
     machine parameters (bias)  &  - &  -  & - \\ 
     machine parameters (non-linearity)  &  $<< $ 1\% & $<< $ 1\%  & $<<$ 1\% \\ 
     experimental (bias) &  $<$ 0.005 & $<$ 0.005  & $<$ 0.005 \\ 
     exp. acceptance (non-linearity) & +3\%  & +5\% & +10\%  \\         
     exp. reconstruction (non-linearity) & -5\% & -5\%  & -15\% \\
     theory (bias)     &     $<< $ 0.001   & $<< $ 0.001 & $<< $ 0.001 \\  
     theory (non-linearity)     &   $\pm$ 2\%    & $\pm$ 0.9\% & $\pm$ 0.5\%  \\ \hline
    \end{tabular}
    \label{tab:systematics}
  \end{center} 
\end{table}

The estimates of the systematic uncertainties on $\mathcal{A}^{Re}$ for several centre-of-mass energies are presented in 
Table~\ref{tab:systematics}. Our study has not found any sources of 
systematic uncertainty that yield a spurious asymmetry
when the true asymmetry is zero. 
Upper limits on a systematic bias in $\mathcal{A}^{Re}$
are given in the table with the label ``(bias)''. 
Several sources can however enhance or dilute a non-zero true asymmetry. These are indicated as the expected relative
modification of the asymmetry, with the label ``(non-linearity)''. Of course, these effects can be corrected to a good
extent using Monte Carlo simulation. The selection bias can moreover be reduced by comparing the measured and predicted
results in an appropriate fiducial region.

\section{Prospects for CP-violating form factors}
\label{sec:precision}

The prospects for a measurement of the top-quark form factors  $F_{2A}^{\gamma,Z}$ 
 are presented in Table~\ref{tab:cpv_values}. Rows two and three of the table show
the result of our simulations described in the preceding sections for a 380~\gev{} stage of 
the Compact Linear Collider CLIC and the initial 500~\gev{} run at the 
International Linear Collider. In both cases an integrated luminosity of
500~\ifb{} is assumed. We find that both projects have a very similar sensitivity to these form factors,
reaching limits of $|F_{2A}^\gamma| <0.01$ for the EDF. Assuming that systematic uncertainties
can be controlled to the required level a luminosity upgrade of either of these machines
may bring a further improvement. The fourth line of Table~\ref{tab:cpv_values} shows the prospects for the nominal ILC scenario, which envisages an integrated luminosity of 4~\iab.

\begin{table}[h!]
        \begin{center}
 	\caption{The expected standard deviations (68\% C.L. limits) of CP-violating form factors derived from 
the statistical precisions on the observables $\mathcal{O}_{\pm}^{Re,Im}$ obtained in this work. 
The results are compared to predictions in the literature, from fast-simulation studies in the context of the TESLA TDR~\cite{AguilarSaavedra:2001rg}, and from studies on the prospects at the (high-luminosity) LHC~\cite{Baur:2004uw,Baur:2005wi,Rontsch:2015una,Schulze:2016qas}, 
 on the potential of a 100~\tev{} proton collider~\cite{Mangano:2016jyj}, and of the LHeC electron-proton
collider~\cite{Bouzas:2013jha}. 
}       
	\begin{tabular}{lccccccc}
		\hline
collider	&  collision & $\sqrt{s}$ & $L_{int}$  & ${\rm Re}~F_{2A}^{\gamma}$& ${\rm Re}~F_{2A}^{Z}$  &  ${\rm Im}~F_{2A}^{\gamma}$ & ${\rm Im}~F_{2A}^{Z}$ \\ 
	&            & [TeV]      & [\iab]    & & & & \\

  \hline
\multicolumn{8}{l}{{\bf Prospects derived in this study:}} \\  
CLIC initial & $e^+e^-$ & 0.38 & 0.5 &  0.015 & 0.019 & 0.013 & 0.026\\ 
ILC initial & $e^+e^-$ & 0.5 & 0.5 & 0.005 & 0.007 & 0.006 &0.010 \\

ILC nominal   & $e^+e^-$ & 0.5 & 4               &  0.002 & 0.003  & 0.002 &  0.004  \\ \hline  

CLIC (parton-level) &  $e^+e^-$ & 3 & 3 &  0.003 & 0.003  & 0.005 &  0.009  \\ \hline 

               \multicolumn{8}{l}{\bf Previous studies for lepton colliders:} \\ 
 TESLA (Aguilar et al.~\cite{AguilarSaavedra:2001rg}) & $e^+e^-$ & 0.5 & 0.3 & 0.007&0.008&0.008& 0.010\\  \hline
\multicolumn{8}{l}{\bf Prospects for hadron colliders:} \\ 
HL-LHC (Baur~et~al.~\cite{Baur:2004uw,Baur:2005wi}) & $pp$ & 14 &  3  & 0.12&0.25&0.12& 0.25\\ 
HL-LHC (R\"ontsch \& Schulze~\cite{Rontsch:2015una}) & $pp$ & 14 &  3  & -   & 0.16 & - & - \\
 FCChh (Mangano et al.~\cite{Mangano:2016jyj}) & $pp$ & 100 & 3 &  -   &  0.04    & -  & - \\
LHeC (Bouzas et al.~\cite{Bouzas:2013jha})  & $ep$ & - & 0.1 &  0.1   &    -    &   -   & - \\
\hline
	\end{tabular}
	\label{tab:cpv_values}
	\end{center}
	\end{table}

The prospects for these measurements at a multi-\tev{} electron-positron 
collider are listed in the row labelled ``CLIC3000'' of 
Table~\ref{tab:cpv_values}.
The sensitivity of the  CP-odd observables studied in this paper 
to $F_{2A}^{\gamma,Z}$ increases, for $\sqrt{s}\gg 2 m_t$, approximately 
linearly with the centre-of-mass energy. On the other hand the cross 
section for $\ttbar$ production via $s$-channel $Z/\gamma^*$-boson exchange
decreases as $1/s$. At linear colliders this is partly compensated by the 
higher luminosity at high energy: typically the instantaneous luminosity 
increases linearly with $\sqrt{s}$. 
All in all, for the 3~\tev{} stage of CLIC the precision is
expected to be significantly higher than for the initial stage 
at $\sqrt{s}=$ 380~\gev.

We recall here that the two-Higgs-doublet extensions of the three-generation 
standard model investigated in Section~\ref{sec:cpvff} give rise to sizeable 
form factors predominantly at centre-of-mass energies close to the 
$\ttbar{}$ production threshold. However, CP-violating new physics models 
 with new heavy particles are conceivable that lead to enhancements of the 
CP-violating  top-quark form factors $F_{2A}^{\gamma,Z}$ in the TeV energy range.

The next row in Table~\ref{tab:cpv_values} lists the results given in 
the TESLA Technical Design Report~\cite{AguilarSaavedra:2001rg}.
The results of our full simulation analysis are in agreement with the 
expectations of this parton-level study, once differences in the 
assumptions on polarization and integrated luminosity are taken
into account.

\begin{figure}[h!]
\centering
\includegraphics[width=0.7\textwidth]{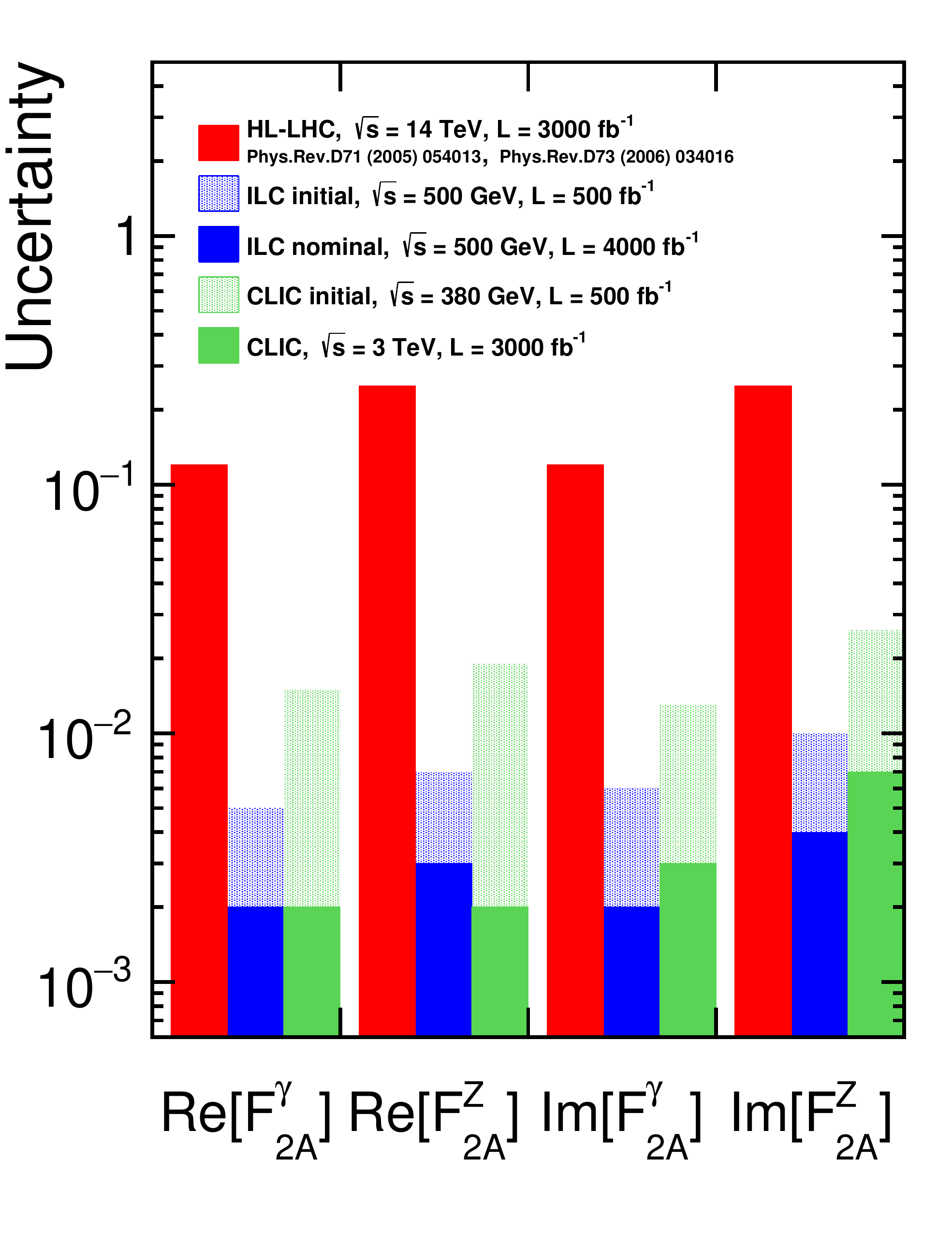} 
\caption{\small{Graphical comparison of 68\% C.L. limits on CP-violating form factors expected 
at the LHC \cite{Baur:2004uw,Baur:2005wi}, and at the ILC and CLIC (this work). 
 The LHC simulations assume an integrated luminosity of $\mathcal{L} = 3000 {\rm fb}^{-1}$ at 14 TeV. 
  For the ILC we assume an initial $\mathcal{L} = 500 {\rm fb}^{-1}$ at 500 GeV and a beam polarization $P_{e^-}=\pm0.8, P_{e^+}=\mp0.3$. The nominal scenario envisages an integrated luminosity of 4~${\rm ab}^{-1}$.
  For CLIC we assume $\mathcal{L} =$ 500~${\rm fb}^{-1}$ at 380 GeV for the initial stage and $\mathcal{L} =$ 3000~${\rm ab}^{-1}$ at 3~\tev{} for the high-energy stage. The electron beam polarization is $P_{e^-}=\pm 0.8$ and no positron polarization is envisaged.}}
\label{fig:Manhattan_plot_nacho_CPV_wo_TESLA_nacho}
\end{figure}

\subsection{Prospects at hadron colliders}

A complete study of measurement prospects on $F_{2A}^{\gamma,Z}$  in the 
associated production 
of top-quark pairs and gauge bosons,   $t\bar{t}Z$ and $t\bar{t}\gamma$, at hadron colliders  was made in 
Refs.~\cite{Baur:2004uw,Baur:2005wi}. The constraints on the
four CP-violating form factors are listed in Table~\ref{tab:cpv_values}
 under the header ``prospects for
hadron colliders''. These results are compared
to our results for the initial ILC and CLIC stages in 
Figure~\ref{fig:Manhattan_plot_nacho_CPV_wo_TESLA_nacho}. Clearly, the
measurements at hadron colliders are expected to be considerably 
less precise than those that can be made at lepton colliders, 
even after completion of the full LHC programme 
including the planned luminosity upgrade.

Furthermore, Table~\ref{tab:cpv_values} summarizes the results of more 
recent studies of the potential of hadron colliders. 
The chirality-flipping terms proportional to $\sigma_{\mu\nu}$ in the 
effective Lagrangian
 used in  Ref.~\cite{Rontsch:2015una} 
(cf. also Refs.~\cite{Schulze:2016qas,Mangano:2016jyj}) 
differ by a factor $2 m_t / m_Z \sim 4$ from our convention defined in 
Eq.~\eqref{eq:vtxvtt}.
Thus the form factors $F_{2A}$  used in this paper are related 
to the couplings $C_{2A}$   of Ref.~\cite{Rontsch:2015una}  
by $F_{2A} = C_{2A}  2 m_{t} / m_Z$. The 95\% C.L. limits on $C_{2V/A}$ given in 
Refs.~\cite{Rontsch:2015una,Schulze:2016qas,Mangano:2016jyj} 
are translated into 68\% C.L. limits on $F_{2V/A}$ to facilitate comparison.

The ultimate prospects of the LHC and the luminosity upgrade depend 
crucially on the control of systematic uncertainties. Ref.~\cite{Rontsch:2015una} 
finds a theory uncertainty of 15\% on the 
total cross section calculated at NLO precision, leading to a 20-40\% 
improvement of the constraint on   ${\rm Re}~F_{2A}^{Z}$ obtained at LO. 
Ref.~\cite{Schulze:2016qas} shows that cross section 
ratios $\sigma_{t\bar{t}Z}/\sigma_{t\bar{t}}$ and $\sigma_{t\bar{t}\gamma}/\sigma_{t\bar{t}}$  
may be calculated to approximately 3\% precision. The HL-LHC and FCChh 
prospects from Ref.~\cite{Mangano:2016jyj} listed 
in Tab.~\ref{tab:cpv_values} assume a systematic uncertainty of 
15\% and 5\%, respectively. 

A lepton-proton collider such as the LHeC~\cite{AbelleiraFernandez:2012cc} 
can provide constraints on anomalous top-quark electroweak couplings 
through measurements of the single top production 
rate ($ep \rightarrow \nu t X$) and the $\ttbar$ photo-production 
rate \cite{Bouzas:2013jha}. These measurements constrain the combination
of the CP-conserving and CP-violating form factors of the top-quark 
interaction with the photon, i.e., on  $F_{2V}^\gamma$ and $F_{2A}^\gamma$ 
in the notation of Section~\ref{sec:anomalous}. Assuming a large 
integrated luminosity (100~\ifb{}) of energetic $ep$ collisions 
($E_{p} =$ 7~\tev, $E_e =$ 140~\gev), Ref.~\cite{Bouzas:2013jha} derives
the expected limit on $F_{2A}^{\gamma}$ that is listed in the last 
row of Tab.~\ref{tab:cpv_values}.

\subsection{Comparison to indirect constraints}

Direct experimental bounds on CP-violating contributions to the $t\bar{t}Z$ 
and $t\bar{t}\gamma$ vertices are not 
available. However, with mild assumptions measurements that yield 
information about the $Wtb$ vertex can be recast into
limits on the form factors of the $\gamma t \bar{t}$ and $Z t \bar{t}$ 
interactions. 
In a dimension-six effective-operator framework based on the SM gauge 
symmetry~\cite{AguilarSaavedra:2008zc,Grzadkowski:2010es,Zhang:2010dr} 
the operator $O_{tW}$ (with Wilson coefficient $C_{tW}$) generates an anomalous
chirality-flipping coupling 
$g_R$ of the $W$-boson (cf. Section~\ref{suse:twb}) and non-zero values 
for the real part of the $F_{2A}^{\gamma,Z}$ form factors in 
$\epem \rightarrow \ttbar$ production. We use this approach to convert 
constraints from measurements 
of the W-helicity fractions in top-quark decay~\cite{Birman:2016jhg,Chatrchyan:2013jna,Aad:2012ky}, 
of the single top production cross sections, and from studies of the polarization of 
$W$-boson in t-channel single-top 
production~\cite{Aaboud:2017aqp,Aad:2015yem} into constraints on  $F_{2A}^{\gamma,Z}$.

Ref.~\cite{Birman:2016jhg} presents a combined fit to $W$-boson helicity fractions and single top 
production cross sections measured at the LHC, resulting in a 95\% C.L. limit of 
${\rm Im} {g_R} \in [-0.30 , 0.31]$, where $g_R$ is one of the two 
chirality-flipping form factors in the $t\to Wb$ decay amplitude, see
Section~\ref{suse:twb}.
We translate this result  into a bound on $F_{2A}^{\gamma,Z}$. First we use the following expression 
from Ref.~\cite{AguilarSaavedra:2008zc} in order to relate $g_R$ to the Wilson coefficient $C_{tW}$ 
of the effective (dimension-6) operator $O_{tW}$:
\begin{equation}
g_{R} = \sqrt{2}  C_{tW} \frac{v^2}{\Lambda^2} \, .
\end{equation}
The result of Ref.~\cite{Birman:2016jhg} can then be
converted into an allowed band for ${\rm Re}~F_{2A}^\gamma$ and  ${\rm Re}~F_{2A}^Z$ using the 
following relations\footnote{As a cross-check the relations between form factors 
and Wilson coefficients in 
Equations~\eqref{eq:relnch1} and \eqref{eq:relnch2} have been verified
using a MadGraph~\cite{Alwall:2014hca} UFO model of the dimension-6 operators 
that affect the top-quark electro-weak vertices. The basis of the model
is presented in Ref.~\cite{Bylund:2016phk}. More recent additions, in 
particular the extension to the CP-violating imaginary parts of the
coefficients will be reported in a future publication~\cite{durieux2017}. 
With this setup and conversion relations we are able to reproduce several
key results of Refs.\cite{AguilarSaavedra:2008zc} and~\cite{Rontsch:2015una}.}:
\begin{equation}
{\rm Re}~F_{2A}^Z = \sqrt{2} \left( \frac{4 m_t^2 }{\Lambda^2 s_W c_W}\right) {\rm Im}[c_W^2 C_{tW} - s_W^2 C_{tB}] 
\label{eq:relnch1}
\end{equation}
and
\begin{equation}
{\rm Re}~F_{2A}^{\gamma} = \sqrt{2} \left( \frac{4 m_t^2 }{\Lambda^2}\right) {\rm Im}[C_{tW} + C_{tB}] \, .
\label{eq:relnch2}
\end{equation}
ATLAS has recently released two measurements of the decay of polarized top quarks in t-channel 
single-top production~\cite{Aaboud:2017aqp,Aad:2015yem} and presented the 95\% C.L. 
limit: ${\rm Im} (g_R/V_L) \in  [-0.18, 0.06]$.
Setting $V_L = V_{tb} \sim $ 1 this leads to a slightly tighter limit on the CP-violating 
dipole operators.
The bands corresponding to both limits are drawn in Fig.~\ref{fig:summaryplot}, where the
prospects listed Table~\ref{tab:cpv_values} are also shown for comparison. 

\begin{figure}[htb!]
	{\centering
	\includegraphics[width=0.89\textwidth]{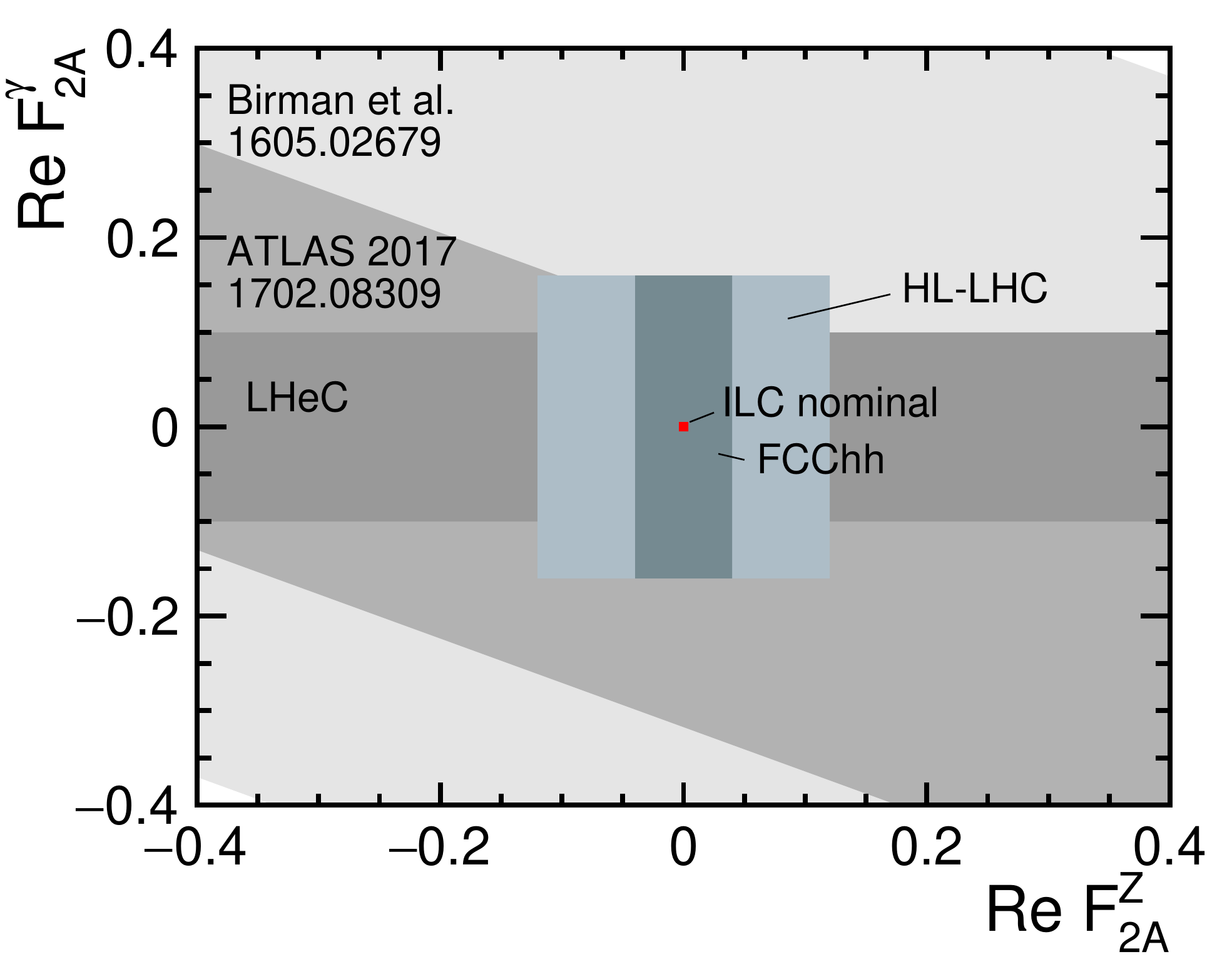}
	\caption{The 68\% C.L. limits on ${\rm Re}~F_{2A}^{Z}$ and ${\rm Re}~F_{2A}^{\gamma}$ 
derived from measurements of the $Wtb$ vertex performed by ATLAS and CMS during run I of the 
LHC. This interpretation assumes that there is a relation between the imaginary part of the
anomalous chirality-flipping coupling $g_R$ that affects the $tWb$-vertex and the real part 
of the form factors $F_{2A}^{\gamma,Z}$ measured in $e^+e^- \rightarrow t\bar{t}$ production, 
as is generally the case in an effective-operator interpretation. The prospects of future 
colliders are indicated for comparison. }
	\label{fig:summaryplot}
	}
\end{figure}



Further indirect bounds can be extracted from data at lower energies. 
Ref.~\cite{deBlas:2015aea} used electroweak precision data to derive 
constraints on top-quark electroweak couplings, but CP-violating operators 
were not taken into account. 
Using in addition experimental upper bounds on the electric dipole moments 
of the neutron and atoms/molecules a powerful indirect constraint was 
derived in Ref.~\cite{Cirigliano:2016njn,Cirigliano:2016nyn} on the static 
moment $F_{2A}^\gamma$ of the top quark.







\section{Conclusions}
\label{sec:conclusions}

CP violation in the top-quark sector is relatively unconstrained by direct 
measurements. While the Standard Model predicts very small effects, which 
are beyond the sensitivity
of current and future colliders, sizeable effects
may occur within well-motivated extensions of the SM.
We have updated, within the type-II two-Higgs-doublet model and the MSSM, the potential magnitude of CP violation in 
the top-quark sector, taking into account constraints of LHC measurements. 
The  CP-violating top-quark form factors $F_{2A}^{\gamma,Z}$ whose static limits are the electric and weak dipole
moments of the top quark can be as large as 0.01 in 
magnitude in a viable 2HDM.  

We have investigated the prospects of detecting CP violation in $t{\bar t}$ 
production at a future $e^+e^-$ collider.
 The top spin-momentum correlations proposed in 
Ref.~\cite{bib:cpvbernreuther2}  for $t{\bar t}$ decay to lepton plus jets 
final states were evaluated
with a full simulation of polarized electron and positron beams 
including a detailed model of the detector response. 
Biases due to the selection and migrations in the distributions 
of observables ${\cal O}_{\pm}^{Re}$ and  ${\cal O}_{\pm}^{Im}$
due to ambiguities in the reconstruction of the 
top-quark candidates were found to cancel in the 
CP asymmetries ${\cal A}^{Re}$ and ${\cal A}^{Im}$ defined
 in Eqs.~\eqref{eq:ARe}, \eqref{eq:AIm}. 
We expect therefore that these asymmetries, which are sensitive
to the CP-violating top-quark form factors  $F_{2A}^{\gamma,Z}$, 
are robust against such effects
and can be measured with good control over experimental 
and theoretical systematic uncertainties. Thus, our results 
validate the findings of an earlier parton-level 
study~\cite{AguilarSaavedra:2001rg} for the TESLA collider.

Measurements of these top spin-momentum correlations at a future lepton 
collider can provide a tight constraint on CP violation in the top-quark 
sector. The 68\% C.L. limits on the magnitudes of the form factors 
${\rm Re} F_{2A}^{\gamma,Z}$ and  ${\rm Im} F_{2A}^{\gamma,Z}$ 
derived from our analysis of assumed 500~\ifb{} of
data collected at 380~\gev{} or 500~\gev{} are expected to
be better than 0.01. An improvement by a further factor of three
may be achieved in the luminosity upgrade scenario of the ILC or
in the high-energy stage of CLIC.
These prospects constitute an improvement by two orders of
magnitude over the existing indirect limits.
With this precision, a linear 
collider can probe the level of CP violation in the top-quark sector
predicted by a viable 2HDM model of Higgs-boson induced CP violation. 

A comparison with the expectations for hadron colliders, as derived 
in Refs.~\cite{Baur:2004uw,Baur:2005wi,Rontsch:2015una,Schulze:2016qas,Mangano:2016jyj}, shows that the sensitivity of a future $e^+e^-$ collider to 
CP-violating dipole form factors is very competitive.
The constraints on form factors represent an order of magnitude improvement 
of the limits expected after the complete LHC programme,
including the planned luminosity upgrade.
The potential even exceeds that of a 100~\tev{} hadron collider,
such as the FCChh.

\section*{Acknowledgements}

We would like to thank our colleagues in the ILD and CLICdp groups. 
The study in Section~\ref{sec:CPILC} was carried out in the framework 
of the ILD detector concept, that of Section~\ref{sec:CPCLIC} in the CLICdp 
collaboration. We gratefully acknowledge in particular the LCC
generator group and the core software group which developed the 
simulation framework and produced the Monte Carlo simulation samples 
used in this study. We thank several members of both collaboration for
a careful review of the manuscript and many helpful suggestions to improve
the paper. This work benefits 
from services provided by the ILC Virtual Organisation, supported by the 
national resource providers of the EGI Federation. This research was 
done using resources provided by the Open Science Grid. 
 The authors at IFIC (UVEG/CSIC) are supported under grant MINEICO/FEDER-UE, 
FPA2015-65652-C4-3-R.
 L. Chen is supported by a scholarship from the China Scholarship Council (CSC).


\clearpage

 \section*{Appendix: coefficients for ${\mathcal A}^{Re}$ and  ${\mathcal A}^{Im}$ }
 \label{sec:appendix}
 Here we give formulas for the  coefficients $c_{\gamma}(s), c_{Z}(s)$ and ${\tilde c}_{\gamma}(s), {\tilde c}_{Z}(s)$ that
  determine the CP asymmetries \eqref{eq:ARe} and \eqref{eq:AIm}, respectively. They can be represented as ratios
  \begin{align}
  c_{\gamma}(s) = \frac{N_{\gamma}(s)}{D(s)} \, , \quad  c_{Z}(s) = \frac{N_{Z}(s)}{D(s)} \, , \quad
  {\tilde c}_{\gamma}(s) = \frac{{\widetilde N}_{\gamma}(s)}{D(s)}\, , \quad  {\tilde c}_{Z}(s) = \frac{{\widetilde N}_{Z}(s)}{D(s)}\, . 
  \end{align}
 We compute the matrix elements for the lepton plus jets finals states at 
tree level, use the narrow width approximation for the intermediate $t$ 
and ${\bar t}$, and integrate over the full phase space. Moreover, we neglect 
the width in the $Z$-boson propagator, since we are sufficiently far away 
from the $Z$ peak and we work to lowest order in the electroweak couplings. 
With the conventions defined in Eq.~\eqref{eq:vtxvtt} and~\eqref{eq:ffactors} 
we obtain:
\begin{align}
N_{\gamma}(s) = & \;  \displaystyle{ -\frac{4 \beta_t \sqrt{s} }{3 m_t} } (s-m_Z^2) v_e^{\gamma} \left\{ (1 - P_{-} P_{+}) s a_e^{Z} v_t^{Z}  \right. \nonumber \\
  & \left. - \, (P_{-} - P_{+})  [(m_Z^2 -s)v_e^{\gamma} v_t^{\gamma}  - s v_e^{Z} v_t^{Z}]\right\} \, ,
  \label{eq:numgam}
\end{align}
\begin{align}
 N_{Z}(s) = & \; \displaystyle{-\frac{4 \beta_t s^{3/2}}{ 3 m_t} } \left\{ (P_{-} - P_{+}) s (a_e^{Z})^2 v_t^{Z} + 
    (P_{-} - P_{+}) v_e^{Z} [(s-m_Z^2) v_e^{\gamma} v_t^{\gamma}  + s v_e^{Z} v_t^{Z}] \right. \nonumber \\
   & \left. - \,  (1 - P_{-} P_{+}) a_e^{Z}  [(m_Z^2 - s) v_e^{\gamma} v_t^{\gamma}  - 2 s v_e^{Z} v_t^{Z}]\right\} \, ,
   \label{eq:numZ}
   \end{align}
 \begin{align}
 {\widetilde N}_{\gamma}(s)  = &  \; \displaystyle{\frac{2 \beta_t }{15 m_t^2} } (16 m_t^2 + s) (s -m_Z^2 ) v_e^{\gamma} \left\{ (P_{-} - P_{+}) s a_e^{Z} v_t^{Z} \right.  \nonumber \\
    & \left.    - (1 - P_{-} P_{+}) \left[(m_Z^2 -s) v_e^{\gamma} v_t^{\gamma} -  s v_e^{Z} v_t^{Z} \right]\right\}  \, ,
   \label{eq:numtigam}
   \end{align}
\begin{align}
{\widetilde N}_{Z}(s)  = & \;  \displaystyle{\frac{2 \beta_t s (16 m_t^2 +  s)}{15 m_t^2}} \left\{  (1 - P_{-} P_{+}) s  (a_e^{Z})^2 v_t^{Z}
 +  (1 -  P_{-} P_{+})  v_e^{Z}  [(s-m_Z^2) v_e^{\gamma} v_t^{\gamma} 
       + s v_e^{Z} v_t^{Z}]    \right.  \nonumber \\
   & \left.    - \, (P_{-} - P_{+}) a_e^{Z} [(m_Z^2 - s)v_e^{\gamma} v_t^{\gamma} -  2 s v_e^{Z} v_t^{Z}]\right\}   \, ,
    \label{eq:numtiZ}
\end{align}
\begin{align}
D = & \; \displaystyle{\frac{4}{s} } \left\{  (1 - P_{-} P_{+}) s^2 (s - 4 m_t^2 ) (v_e^{Z} a_t^{Z})^2  \right. \nonumber \\
      & \left.  + \,  (1 - P_{-} P_{+}) (s + 2 m_t^2) [(s -m_Z^2) v_e^{\gamma} v_t^{\gamma}  + s v_e^{Z} v_t^{Z}]^2 \right. \nonumber \\
      & \left. +   \,   (1 -  P_{-} P_{+}) s^2 (a_e^{Z})^2[(a_t^{Z})^2 (s - 4 m_t^2) + (s + 2 m_t^2 ) (v_t^{Z})^2] \right. \nonumber \\
        & \left.  + \,   2(P_{-} -  P_{+}) s  a_e^{Z}  \left[v_e^{Z} (a_t^{Z})^2 s (s - 4 m_t^2 ) +  v_t^{Z} (s + 2 m_t^2) 
        [(s-m_Z^2) v_e^{\gamma} v_t^{\gamma}  + s v_e^{Z} v_t^{Z}]\right]\right\} \, , \nonumber \\
     \label{eq:denom}   
\end{align}
 where $P_{-} \equiv P_{e^-}$ and   $P_{+} \equiv P_{e^+}$ are the longitudinal polarization degrees of the $e^\mp$ beams, $m_t$ and $m_Z$ denote the mass of the $t$ quark
  and $Z$ boson, respectively, and  $\beta_t= \sqrt{1 - 4 m_t^2/s}$. 
The electroweak couplings of $f= e^-, t$  are 
\begin{equation} \label{eq:EWcouplings}
 v^Z_f=\frac{1}{2 s_W c_W}\left(T_{3f}-2 s_W^2 Q_f \right) \, , \quad a^Z_f = -\frac{1}{2 s_W c_W} T_{3f} \, , \quad v_f^\gamma = Q_f \, , \quad a_f^\gamma= 0 \, , 
\end{equation}
 where $T_{3f}$ is the third component of the weak isospin of $f$, $Q_f$ is the electric
  charge of $f$ in units of $e>0$, and $s_W, c_W$ are the sine and cosine of the weak mixing angle $\theta_W$. 
 We have neglected terms bilinear in the CP-violating form factors in the computation of the denominator $D$, because we know a posteriori
  that $|F_{2A}^{\gamma,Z}|$ must be significantly smaller than one. 
  
%



\bibliographystyle{utphysmod}

\begin{footnotesize}
\bibliography{CPV-refs_v7}

\end{footnotesize}

\end{document}